\documentclass[preprint,12pt]{elsarticle}

\usepackage{amssymb}
\usepackage{amsthm}
\usepackage[top=2.5cm,bottom=2.5cm,left=2.5cm,right=2.5cm]{geometry}
\usepackage[noend,ruled]{algorithm2e}
\usepackage{siunitx}
\usepackage{url}
\usepackage{amsmath}
\usepackage[short]{optidef}
\usepackage{nomencl}
\usepackage{etoolbox}
\usepackage{framed}
\usepackage{multicol}
\usepackage{longtable}
\usepackage{hyperref}
\makenomenclature
\newcommand\blfootnote[1]{%
  \begingroup
  \renewcommand\thefootnote{}\footnote{#1}%
  \addtocounter{footnote}{-1}%
  \endgroup
}

\renewcommand\nomgroup[1]{%
  \item[\bfseries
  \ifstrequal{#1}{A}{Variables}{%
  \ifstrequal{#1}{B}{Sets}{%
  \ifstrequal{#1}{C}{Operators}{%
  \ifstrequal{#1}{D}{Indices}{
  \ifstrequal{#1}{E}{Superscripts}{}
  }}}}%
]}

\renewcommand{\nompreamble}{\begin{multicols}{2}\scriptsize}
\renewcommand{\nompostamble}{\end{multicols}}

\makeatletter
\def\ps@pprintTitle{%
 \let\@oddhead\@empty
 \let\@evenhead\@empty
 \def\@oddfoot{}%
 \let\@evenfoot\@oddfoot}
\makeatother

\begin{document}

\begin{frontmatter}
\title{Comparing Approaches to Distributed Control of Fluid Systems based on Multi-Agent Systems}

\author[label1]{\texorpdfstring{Kevin T. Logan\fnref{label2}}}
\author{\texorpdfstring{J. Marius St{\"u}rmer\fnref{label2}}}
\author[label1]{Tim M. M{\"u}ller}
\author[label1]{\texorpdfstring{Peter F. Pelz\corref{cor1}}{Peter F. Pelz}}
\cortext[cor1]{Corresponding Author}
\ead{peter.pelz@fst.tu-darmstadt.de}
\ead[url]{https://www.fst.tu-darmstadt.de/fachgebiet/index.en.jsp}
\address[label1]{Chair of Fluid Systems, Technische Universit{\"a}t Darmstadt, \mbox{Otto-Berndt-Stra{\ss}e 2, 64287 Darmstadt, Germany}}
\fntext[label2]{These authors contributed equally to the presented work and are both to be considered first authors.}

\begin{abstract}
Conventional control of fluid systems does not consider system-wide knowledge for optimising energy efficient operation.
Distributed control of fluid systems combines reliable local control of components while using system-wide cooperation to ensure energy efficient operation.
The presented work compares three approaches to distributed control based on multi-agent systems,
distributed model predictive control (DMPC), multi-agent deep reinforcement learning (MADRL) and market mechanism design.
These approaches were applied to a generic fluid system and evaluated with regard to functionality,
energy efficient operation, modeling effort, reliability in the face of disruptions, and transparency of control decisions.
All approaches were shown to fulfil the functionality, though a trade-off between functional quality and energy efficiency was identified.
Increased modeling effort was shown to improve the performance slightly
while a strong interdependence of information caused by excessive information sharing has proven to be disadvantageous.
DMPC and partially observable MADRL were less sensitive to disruptions than market mechanism.
In conclusion, agent-based control of fluid systems achieves greater energy efficiency than conventional methods,
with values similar to centralized optimal control and thus represent a viable design approach of fluid system control.
\end{abstract}


\end{frontmatter}
\blfootnote{\textit{Abbreviations:}
DMPC, distributed model predictive control;
FIPA, Foundation of Intelligent Physical Agents;
FMI, Functional Mockup Interface;
FMU, Functional Mockup Unit;
HVAC, Heating, Ventilation, and Air Conditioning;
KIF, Knowledge Interchange Format;
KQML, Knowledge Query and Manipulation Language;
MAAC, multi-agent actor-attention-critic algorithm;
MADRL, multi-agent deep reinforcement learning;
MARL, multi-agent reinforcement learning;
MPC, model predictive control.}

\begin{table*}[tbp]
  \begin{framed}
\nomenclature[A, 01]{\(A\)}{Cardinality of \(\mathcal{A}\)}
\nomenclature[A, 02]{\(A'\)}{Cross section}
\nomenclature[A, 03]{\(a\)}{Action}
\nomenclature[A, 04]{\(B\)}{Budget}
\nomenclature[A, 05]{\(b\)}{Price in offer calculation}
\nomenclature[A, 06]{\(\hat{C}\)}{Cost function of prediction horizon}
\nomenclature[A, 07]{\(c\)}{Cost function}
\nomenclature[A, 08]{\(\mathcal{D}\)}{Experience replay buffer}
\nomenclature[A, 09]{\(d_{\mathrm{power}}\)}{Power conversion constant}
\nomenclature[A, 10]{\(g\)}{Steady state model}
\nomenclature[A, 11]{\(J\)}{Policy}
\nomenclature[A, 12]{\(k\)}{Mapping function of volume flow controller}
\nomenclature[A, 13]{\(\mathcal{L}\)}{Loss function}
\nomenclature[A, 14]{\(L\)}{Prediction horizon}
\nomenclature[A, 15]{\(l\)}{Pressure loss parameter}
\nomenclature[A, 16]{\(M\)}{Number of discrete values for manipulated variables}
\nomenclature[A, 17]{\(N_{\mathrm{offers}}\)}{Cardinality of \(O\)}
\nomenclature[A, 18]{\(n\)}{Pump speed}
\nomenclature[A, 19]{\(o\)}{Observation}
\nomenclature[A, 20]{\(P\)}{Power}
\nomenclature[A, 21]{\(p\)}{Pressure}
\nomenclature[A, 22]{\(Q\)}{Volume flow}
\nomenclature[A, 23]{\(r\)}{Reward}
\nomenclature[A, 24]{\(S\)}{Safety factor}
\nomenclature[A, 25]{\(s\)}{State}
\nomenclature[A, 26]{\(u\)}{Manipulated vaiable}
\nomenclature[A, 27]{\(V\)}{Objective function of (D)MPC}
\nomenclature[A, 28]{\(v\)}{Valve opening}
\nomenclature[A, 29]{\(w\)}{Flow velocity}
\nomenclature[A, 30]{\(x\)}{System state}
\nomenclature[A, 31]{\(y\)}{System output}
\nomenclature[A, 32]{\(\alpha\)}{Coefficient of pressure characteristic of pump}
\nomenclature[A, 33]{\(\beta\)}{Coefficient of power characteristic of pump}
\nomenclature[A, 34]{\(\gamma\)}{Discount factor}
\nomenclature[A, 35]{\(\delta\)}{Restriction range of manipulated variable}
\nomenclature[A, 36]{\(\zeta\)}{Pressure loss coefficient}
\nomenclature[A, 37]{\(\theta\)}{Policy function weights}
\nomenclature[A, 38]{\(\lambda\)}{Cost weighting factor}
\nomenclature[A, 39]{\(\pi\)}{Policy function}
\nomenclature[A, 40]{\(\varrho\)}{Density}
\nomenclature[A, 41]{\(\rho\)}{Agent cost weighting factor}
\nomenclature[A, 42]{\(\varphi\)}{Temperature parameter}
\nomenclature[A, 43]{\(\chi\)}{Action value function}
\nomenclature[A, 44]{\(\psi\)}{Value function weights}

\nomenclature[B]{\(\mathcal{A}\)}{Set of agents}
\nomenclature[B]{\(\mathcal{O}\)}{Set of observations}
\nomenclature[B]{\(O\)}{Set of offers}
\nomenclature[B]{\(\mathcal{P}\)}{Set of pump agents}
\nomenclature[B]{\(\mathbb{P}\)}{Manipulated variable range of pump controller}
\nomenclature[B]{\(\mathcal{S}\)}{Set of states}
\nomenclature[B]{\(\mathcal{T}\)}{Set of time steps}
\nomenclature[B]{\(\mathcal{V}\)}{Set of valve agents}
\nomenclature[B]{\(\mathbb{V}\)}{Manipulated variable range of valve controller}

\nomenclature[C]{\(\mathbb{E}\)}{Expectation Operator}
\nomenclature[C]{\(\nabla\)}{Gradient}

\nomenclature[D]{\(\mathrm{actual}\)}{Actual volumen flow at valves}
\nomenclature[D]{\(\mathrm{demand}\)}{Required volume flow at valves}
\nomenclature[D]{\(\mathrm{est}\)}{Estimated}
\nomenclature[D]{\(\mathrm{global}\)}{Considering all agents}
\nomenclature[D]{\(i\)}{Running index denoting agents}
\nomenclature[D]{\(\mathrm{in}\)}{Inlet}
\nomenclature[D]{\(\mathrm{ind}\)}{Individual}
\nomenclature[D]{\(j\)}{Running index denoting predicted steps}
\nomenclature[D]{\(k\)}{Running index denoting control steps}
\nomenclature[D]{\(\mathrm{max}\)}{Maximum}
\nomenclature[D]{\(\mathrm{min}\)}{Minimum}
\nomenclature[D]{\(\mathrm{out}\)}{Outlet}
\nomenclature[D]{\(\mathrm{set}\)}{Set point}
\nomenclature[D]{\(t\)}{Time stept}
\nomenclature[D]{\(\mathrm{up}\)}{Upstream}
\nomenclature[D]{\(z\)}{Running index denoting agents}

\nomenclature[E]{\(\hat{\square}\)}{Predicted}
\nomenclature[E]{\(\bar{\square}\)}{Updated weights}
\nomenclature[E]{\(\square^T\)}{Transpose}
\nomenclature[E]{\(\square'\)}{Value at next time step}
    \printnomenclature
  \end{framed}
\end{table*}

\section{Introduction}
\label{sec:intro}

Fluid systems are necessary to provide drinking water, to cool chemical processes, to dispose waste water or to perform dosing tasks in food production.
The manufacturers, planners and operators of such systems are confronted with a range of different challenges.
The operation should be uninterrupted and energy-efficient, with minimum investment and personnel costs.
The functionality should always be guaranteed and a change in the system or the boundary conditions should not affect it.
In summary, this results in the five main challenges for fluid systems:

\begin{enumerate}
    \item \emph{Functionality}: The functionality of fluid systems and their control systems is to ensure the specified demand,
                                i.e. a certain pressure or volume flow at specific points in the system.
                                For this purpose, the speed or on/off state of the pumps and the valve positions can be adapted.
                                The functional quality of the system is determined by the deviation of the actually provided volume flow from the control objective.
    \item \emph{Low effort regarding energy consumption}: Pumps and associated fluid systems are responsible for about $8\,\%$ of the total electrical energy consumption \cite{EU.2018, eurostat.2015},
                                                            in industry for up to $25\,\%$ \cite{BMWi.2010}.
                                                            For this reason, the energy efficiency of fluid systems is crucial to ensure the economic efficiency of processes
                                                            and to achieve the sociopolitical goal of reducing $\mathrm{CO}_2$ emissions.
                                                            In operation, up to $20\,\%$ of the energy consumption can still be saved by considering the entire system \cite{Went.2008}.
                                                            Thus, the overall control objective is to fulfill the functionality with as little energy effort as possible.
                                                            In principle, it is favorable if the valves are opened as far as possible to keep the required hydraulic power low.
                                                            However, the demand must be met at all valves.
                                                            In addition, pumps should run at the lowest possible speed and,
                                                            if possible, the pumps with the highest efficiency in the operating point should be active.
    \item \emph{Low effort regarding implementation}: Commissioning requires tuning of the system and design of controllers.
                                                        If there is a change in the system, e.g. due to wear or changed boundary conditions,
                                                        a new tuning in the system is necessary to ensure the functionality and energy efficiency.
                                                        Since fluid systems are usually highly customized, these steps have to be carried out individually,
                                                        which is time-consuming and costly. Thus, when designing fluid systems and their controllers,
                                                        modelling effort should be minimized.
    \item \emph{High availability}: The above examples show that fluid transport is indispensable for many processes.
                                    A malfunction often affects entire production plants or leads to the failure of critical infrastructure,
                                    which is why robust systems with high availability are necessary.
                                    This is generalized as the ability to fulfil the system functionality even in the case where a disruption occurs.
    \item \emph{High acceptability}: Since fluid systems are vital for many social needs and industrial purposes, transparency, traceability and comprehensibility
                                    of control tasks are critical for acceptability of the systems.
                                    Furthermore, the flexibilisation of production, as it is being advanced in the context of Industry 4.0,
                                    demands a high adaptability of technical systems.
                                    Fluid systems are not exempt from this.
                                    Thus, fluid systems will have to react to different and changing boundary conditions in the future.
                                    Fluid system components need to be flexible enough for them to be used in a variety of scenarios.
                                    Thus, portable, flexible and scalable fluid systems and controller designs are an important challenge.
\end{enumerate}

\begin{figure}[h]
    \includegraphics[width=15 cm]{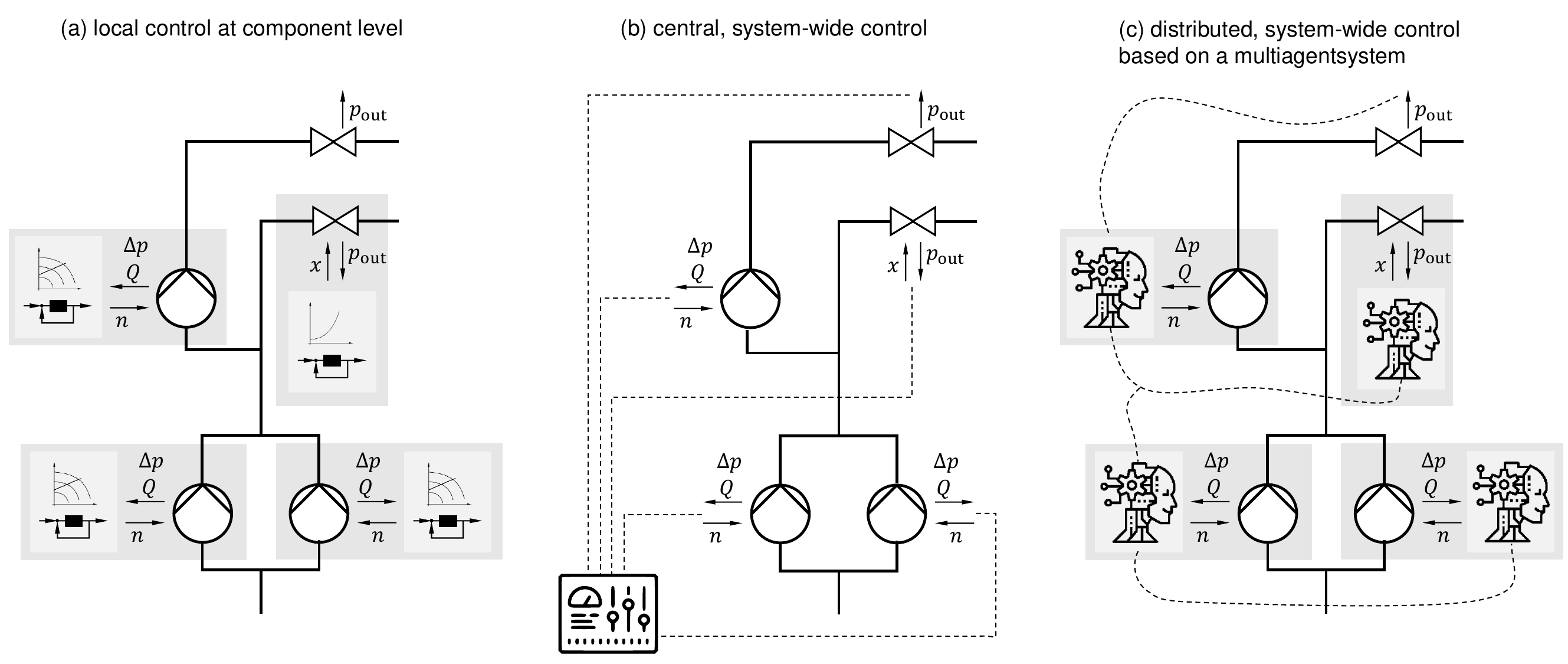}
    \caption{Concepts of three different approaches for controlling a fluid system consisting of pumps and valves. \label{fig:concepts}}
    \end{figure}

To meet the five challenges, an appropriate control architecture and operation strategy are crucial.
In general, three approaches for the control architecture can be distinguished, as shown in Figure \ref{fig:concepts}:
(a) a decentral, local control, (b) a central, system-wide control, and (c) a distributed system-wide control.
Each of these architectures is presented in the following.

In conventional control approaches, the control of the pumps and valves is usually separated, i.e. a local control according to Figure~\ref{fig:concepts} (a).
The underlying concept is that the pumps ensure a certain pressure difference (possibly as a function of the volume flow) and the individual valves at the consumers have a local control for a volume flow, filling level or temperature (e.g. in heat exchangers such as thermostatic valves in a heating system).
The pressure applied by the pumps always has to be high enough to enable the valves to operate and to reach the desired setpoints.
With local control (a), which is also referred to as \textit{decentralized} control, the system is separated into subsystems, each supplied with a local controller.
The subsystems may be constituted only of individual components.
The optimal control objectives are pursued on the local level, while coupling between the subsystems is disregarded~\cite{Pannocchia.2013b}.
On the one hand, this reduces the effort required for tuning and a minimum function is often ensured even in the case of malfunctions.
On the other hand, however, the potential for energy savings given by including information on the system as a whole cannot be fulfilled when components are controlled locally.

In contrast to this, there are system-wide or \textit{centralized} control approaches, which draw on the control inputs from the entire system and calculate the control output for each component to achieve a global optimum for the common objective~\cite{Pannocchia.2013b}.
An example is model-predictive control, in which an optimization is carried out based on a substitute model and thus the optimal control variables are estimated.
As shown in~\cite{Muller.2020}, this can significantly reduce the energy demand.
However, this poses a different challenge -- the high communication requirements due to the merging of all measured and controlled variables and the development, implementation and maintenance of the model.
This is often not worthwhile for highly individualized systems such as a fluid system, which is why these approaches are not very popular in practice.

Ideally, a solution is simple to implement and yet efficient and robust, i.e. combines the advantages of variants (a) and (b).
One such solution is \textit{distributed} control, where local controllers coordinate actions to enable the system as a whole to meet the global control objective, as shown in Figure~\ref{fig:concepts} (c)~\cite{Pannocchia.2013b}.
These local controllers can be modelled as agents which jointly constitute a multi-agent system.
In these systems, each component is assigned a so-called virtual agent, which acts as its control and communication unit.
Each agent has domain-specific knowledge, decision rules and goals.
Furthermore, agents can read sensors and control actuators at the associated component.
The agents can thus perceive and influence the environment, for which they communicate with each other and exchange available knowledge on their respective perceptions and actions.
In the process, the agents make independent and autonomous decisions to perform actions.
The aim is to achieve system-wide targets as a result of the combined effects of these actions without a central control unit.
Conflicting goals are resolved through virtual negotiation between the agents.

Consequently, distributed control architectures based on multi-agent systems promise great potential for application to fluid systems.
However, it is necessary to ascertain the suitability of distributed agent control for addressing the five key challenges presented above.
This entails a systematic study and assessment of the performance of distributed multi-agent systems for control of fluid systems.
For this, both the system-wide rules and procedures as well as the behavior of the individual agents need to be defined.
Furthermore, an appropriately generic fluid system needs to be selected together with a suitable usage scenario.
Finally, benchmarks and assessment criteria need to be defined to gauge to what extent the challenges are addressed by multi-agent systems.
These steps were taken within the scope of the presented work and are detailed in the following.
First, however, the working hypotheses are presented together with the concrete research questions.

\subsection{Research Question}
\label{sec:rq}

When considering distributed control of fluid systems based on multi-agent systems there is a plethora of variations of implementation.
These variations need to be systematically assessed and sorted into a schema determined by their characteristics regarding a set of requirements.

Following from the above, the requirement for distributed multi-agent control of fluid systems is to ensure the functionality.
While this holds for all control designs of fluid systems, the expected behavior of multi-agent control with regard to the challenges $2$ through $4$ differs from that of conventional control.
Three distinct hypotheses on the expected deviation are formulated concerning each of the three challenges.

\textbf{Hypothesis 1:}  Minimum energy consumption is reached when all components of a fluid system are controlled centrally using mathematical global optimization methods. Transitioning from central to distributed agent-based control comes at the cost of increased energy consumption.
Nevertheless, agent-based control still uses optimization methods and consumes less energy than conventional local control methods.
Central optimal control and local control constitute lower and upper boundaries respectively for energy consumption of fluid systems under distributed multi-agent control.

\textbf{Hypothesis 2:} Creating substitute models for highly individual and evolving fluid systems involves prohibitive effort.
Fluid systems can be controlled adequately using control architectures that dispense with manually created substitute models using distributed agents.
This allows for adaptability to changed boundary conditions, as well as transfer of agents between different applications and settings.

\textbf{Hypothesis 3:} In both central and distributed control, the performance of the system's control depends on the communication infrastructure.
Failures in the communication infrastructure adversely affect the performance of central control architectures more severely than distributed control architectures.
Thus, distributed control increases the availability of fluid systems.

Three design approaches for distributed control of fluid systems are selected and examined thoroughly in the light of the requirements addressed above.
These design approaches are (i) distributed model predictive control, (ii) multi-agent deep reinforcement learning, and (iii) market mechanism design.
Furthermore, three concrete research questions guide the design of this study to achieve an assessment and comparison of the three design approaches.
These questions are:

\begin{quote}
    \begin{itemize}
        \item[\textit{(i)}]	\textit{How well do the controllers of each of the design approaches fulfill the functionality and the objective of minimum energy consumption?
        }
        \item[\textit{(ii)}] \textit{What influence does the available information from substitute model and information exchange between agents have with regard to functionality and effort?
        }
        \item[\textit{(iii)}] \textit{How well do the distributed approaches cope with a disruption in the communication between agents in comparison to centralized approaches?}
\end{itemize}
\end{quote}

Focusing on these research questions with the chosen design approaches, the presented work aims to shed light on the hypotheses outlined above.
The approaches were chosen after a thorough examination of the literature.
Relevant previous studies are presented in the following section.

\subsection{Related Work}

Multi-agent systems and agent control of technical systems have been widely studied as approaches to distributed control.
The following section will first give an overview of multi-agent systems applied to technical systems in general.
It will then present various studies investigating multi-agent systems applied to specific technical systems.
Furthermore, the section will present work related to the three approaches to distributed control investigated in the scope of the presented study.

\subsubsection{Agent-based Control of Technical Systems}

Studies presented in the 1980s investigated distributed problem-solving~\cite{Smith.1981, Cammarata.1988}, with communication and interaction between agents playing an important role in early studies of multi-agent systems~\cite{Georgeff.1988, Werner.1989}.
This was further investigated by Jennings~\cite{jennings_1993} and applications of the theory of agent technology in practice discussed by Jennings and Wooldridge~\cite{Jennings.1995, wooldridge_jennings_1995} while approaches to implementing agents in software and programming were also presented by Genesereth and Ketchpel~\cite{Genesereth.1994} as well as Shoam~\cite{Shoham.1994} in the 1990s.
Evans et al. formalized the implementation into a methodology for engineering systems of software agents (cited in~\cite{Bogg.2008}).
Since then, extensive text books have been published introducing multi-agent systems, distributed artificial intelligence, and applications, summarizing the state of the art~\cite{Wooldrige.2002, Vlassis.2007, Kacprzyk.2010}.
Furthermore, guides to assessing suitability and deciding on use of multi-agent systems have been published by Bogg et al.~\cite{Bogg.2008} and Beydoun et al.~\cite{Beydoun.2013}.
A more recent comprehensive survey introducing agents and multi-agent systems as well as discussing in detail applications and challenges was presented by Dorri et al.~\cite{Dorri.2018b}.

Among the applications of multi-agent systems mentioned by Dorri et al. are cities and the built environment, i.e., infrastructure systems and buildings~\cite{Dorri.2018b}.
Negenborn studied these applications to infrastructure, both in transportation networks (Negenborn et al.~\cite{Negenborn.2006}) and power networks~\cite{Negenborn.2007}.
Faced with decentralization connected with the use of renewable energies, power networks are being transformed to smart grids.
How to control these using multi-agent systems was explored by Zimmermann et al.~\cite{Zimmermann.2019}.
Algarvio et al.~\cite{Algarvio.2020} designed a multi-agent system for hydroelectric power plant control in an energy market with high proportion of renewable energies.

Multi-agent system application to control of water distribution networks have also been studied~\cite{Dotsch.2010, Giannetti.2005}.
While not explicitly using multi-agent systems, two studies have investigated distributed control of sewage systems and waste water treatment plants and compared performance of distributed and centralized control systems~\cite{Zhang.2019b, Cembellin.2020}.

Turning from infrastructure systems to buildings, multi-agent systems have been studied extensively in the application to heating, ventilation and air conditioning (HVAC) systems.
Simulations and a field test of multi-agent control of a heating and ventilation system for a commercial building were carried out and compared by Constantin et al.~\cite{Constantin.2016}, while van Pruissen et al.~\cite{vanPruissen.2014} compared multi-agent control of a heating system in a commercial building to conventional, centralized heating system control.
A dissertation by Huber showed that modularized agents could be combined for automated HVAC systems~\cite{Huber.2017}.

Looking beyond functionality, two studies focused on energy cost reduction using multi-agent systems in various configurations for control of HVAC systems~\cite{Wei.2017, Yang.2019b, Yu.2021}.
Azuatalam et al. considered whole-building HVAC agent control with a special focus on-demand response~\cite{Azuatalam.2020}, while Nagarathinam et al. investigated scaling of multi-agent systems in building HVAC with multiple units including water-side control of air-handling-units and distributed agents~\cite{Nagarathinam.2020}.
HVAC systems in buildings demonstrate significant similarities to water distribution systems in buildings.
Nevertheless, building automation and multi-agent control for pumping stations and valves have not been studied hitherto.
A challenge lies in finding an approach that allows studying the performance of various designs of multi-agent systems for fluid system control in a generalized manner.
This allows to transfer promising results to other fluid systems, e.g., industrial cooling circuits, booster stations of high-rise buildings, and urban water distribution systems.

While distributed control using multi-agent systems in general has been applied to technical systems in numerous studies, approaches to designing the multi-agent system can vary considerably.
The presented work compares three such approaches commonly found in the related literature, distributed model predictive control (DMPC) from the domain of control technology, multi-agent deep reinforcement learning (MADRL) from the domain of machine learning, and market mechanism design from the domain of game theory.
Prior works of all three approaches will be presented in the following.

\subsubsection{Distributed Model Predictive Control}

Model predictive control (MPC) has a long history reaching to the 1970s and 1980s with receding horizon feedback control~\cite{Chen.1982, Keerthi.1988, Michalska.1993} and generalized predictive control~\cite{Clarke.1987, Clarke.1987b}.
A survey by Qin and Badgwell~\cite{Qin.2003} not only gives an idea of industrial applications of MPC but also an overview of notable works throughout the history of MPC, and the interested reader is referred to that work for further detail.

An early survey by García et al. describes MPC as using an explicit model that can be identified separately~\cite{Garcia.1989}.
One of the major challenges connected with MPC is to attain that system model.
Various approaches exist, roughly classifiable as analytical and data-driven.

Analytical models require physical modelling of the system and make the inner structure of the system transparent.
One approach is to identify the transfer function of the system with, e.g., a step response or impulse response~\cite{Seborg.2011,Keyser.1988, holkar2010overview}.
Other approaches are to derive a non-linear dynamic model of the system and linearize it for use in MPC~\cite{Bemporad.2009}.

Data-driven system models do not focus on the inner structure of the system, but rather derive a model from input-output-values of a black-box~\cite{Seborg.2011}.
How to derive black-box models from input-output measurements is detailed by Ljung~\cite{Ljung.2001}.
One such approach are ARX models (AutoRegressive models with eXogenous inputs) as they are structurally similar to state space models and the parameters can be identified through convex optimization~\cite{Huusom.2012, Peng.2004, Peng.2007}.
A further black-box approach to deriving a system model is Gaussian Process Regression~\cite{Hewing.2020, Kocijan.2004, Rasmussen.2003, Walder.2008, LazaroGredilla.2010, Snelson.2006}.

Both approaches, physical modelling and models derived from input-output data, have also been combined in the use-case of building modeling~\cite{Hazyuk.2012}.
Finally, an effective approach has been to train neural networks based on empirical input-output data and use these as surrogate system models~\cite{Psichogios.1991, Draeger.1995, Piche.2000}.
This has been done for chemical processes~\cite{Su.1993, Wu.2019} as well as industrial process control~\cite{Kittisupakorn.2009} and automotive control~\cite{Dahunsi.2009}.

It has been argued that applying centralized MPC to large systems can cause problems due to data transmission requirements, computational requirements and considerations of centralized controllers posing a single point of failure~\cite{Yang.2019, Pannocchia.2013b, Stewart.2010b}.
Alternative approaches are decentralization and distribution, where decentralization means dividing a system into subsystems controlled on a local level without strong considerations of coupling whereas distribution entails local controllers of subsystems cooperating for a common, system-wide objective~\cite{Pannocchia.2013b, Stewart.2010b}.
Distributed approaches stem from studying parallel and distributed computing~\cite{Yang.2019,Bertsekas.1997}.
Combining multi-agent systems with distributed model predictive control to solve optimization problems in large systems has been reported~\cite{Vlassis.2007, Negenborn.2006, Negenborn.2007, Negenborn.2009, Stewart.2010b, Pannocchia.2013b,Fioretto.2018, Yang.2019}.

To summarize, MPC is a well-studied control approach for centrally computed optimal control strategies.
DMPC allows to ensure cooperation between local control units of subsystems while still using optimization methods for calculating the control strategy.
Both approaches require a substitute model of the entire system.
The modeling effort may be reduced by applying machine learning methods for deriving this model.

\subsubsection{Multi-Agent Deep Reinforcement Learning}

A different approach has been to avoid modeling the underlying system.
Here, machine learning approaches are implemented directly in the agents controlling the system.
These agents learn based on samples generated through trial and error.

In this approach, reinforcement learning has been used~\cite{LITTMAN1994157, Foerster.2016, Zhang.2021, HernandezLeal.2019}.
This technique is extensively presented in the work of Sutton and Barto~\cite{Sutton.2018}.
Reinforcement learning is a machine learning approach in which the agent is rewarded for executing desirable actions within its environment~\cite{Sutton.2018}.
An overview is given by Zhang et al.~\cite{Zhang.2021} of various algorithms combining multi-agent systems with reinforcement learning based either on Markov games or extensive-form games in cooperative, competitive or mixed settings.
The perception of various agents interacting with one another in a Markov game goes back to Littman~\cite{LITTMAN1994157}.
In a survey, Li presents works showing how reinforcement learning can be extended to deep learning~\cite{Li.2018}.

A further enhancement of general machine learning techniques, deep learning is a technique that requires less human engineering input in designing feature layers.
Instead, it uses generalized machine learning techniques for learning these features from sample data~\cite{LeCun.2015}.
Deep learning is extensively described by Goodfellow et al. in the Deep Learning Book~\cite{Goodfellow.2016}.
Foerster et al. combined deep learning and reinforcement learning for machine learning of communication protocols for multi-agent systems~\cite{Foerster.2016}.
An overview and critical assessment of multi-agent deep reinforcement learning (MADRL) is presented by Hernandez-Leal et al. with regard to current state of the art, recent advances, lessons learned, and challenges for future implementations~\cite{HernandezLeal.2019}.

To make the learning process of agents more efficient, attention weights can be introduced, as shown by Vaswani et al.~\cite{Vaswani.2017}.
A further approach to enhancing the learning process of agents has been to separate them into actors and critics, where actors learn the decision policies and critics learn the value function~\cite{Konda.1999, HernandezLeal.2019}.
This has been applied to multi-agent systems with reinforcement learning with variations~\cite{Lowe.2017, Iqbal.2018}.
Here, Iqbal and Sha used a configuration for decentralized actors while critics learn a global value function subject to attention weights, the multi-agent actor-attention-critic algorithm (MAAC)~\cite{Iqbal.2018}.
Further adaptations have been to decentralize the critic~\cite{Li.2020} as well as giving critics separate hierarchical attention levels for the agent system and the individual agent~\cite{Wang.2020b}.

In machine learning, agent systems are modelled as Markov games.
MADRL is a promising approach for relying on fully data-driven methods for control of fluid systems.
MAAC can be used for an effective training process of agents.

\subsubsection{Market Mechanism Design}

The third approach to distributed control of fluid systems considered in this study is market mechanism design.
Mechanism design is a field of research in the domain of game theory and economics~\cite{Lavi.2012}, yet has also been presented alongside other approaches of distributed artificial intelligence~\cite{Vlassis.2007}.

While game theory is commonly associated with agent systems as a modelling approach in social sciences, it bears similarities to control systems with multiple distributed controllers~\cite{Marden.2018}.
According to Marden and Shamma, the difference in usage in the two disciplines is that the social sciences use it as a "descriptive" method whereas in engineering it is a "design" method~\cite[p.~862f]{Marden.2015}.
They further present some examples where a global objective function serves to ensure a desirable emergent behavior resulting from individual agents' permissible actions in engineering problems, such as synchronization, distributed routing, sensor coverage, wind energy harvesting~\cite{Marden.2012}, vehicle target assignment, content distribution, and ad-hoc networks~\cite{Marden.2015}.
Furthermore, for distributed control using game theoretic approaches, the utility functions of individual agents need to be defined in such a way that the agents' actions contribute to the global objective function~\cite{Marden.2015}.
Finally, when the desirable outcome is defined according to the global objective function, the interaction protocol between the agents needs to be defined in such a way that the agents learn to converge towards that desirable outcome~\cite{Marden.2015}.
The process of designing such a protocol is known as mechanism design~\cite{Vlassis.2007}.

There is a variety of choices for the interaction protocol, which can be roughly categorized as negotiations and auctions~\cite{Wooldrige.2002}.
According to Wooldridge, negotiations are more generally applicable than auctions as they allow reaching a common agreement in a wide range of settings, whereas auctions only consider the problem of allocating goods~\cite[p.~137]{Wooldrige.2002}.
Various agent languages exist for facilitating the required semantic communication between agents during negotiations, e.g., KIF, KQML, one of them being the protocol standardized by the Foundation of Intelligent Physical Agents (FIPA)~\cite{Wooldrige.2002, Bellifemine.2001}.
Bellfemmine et al. implemented this protocol in a Java software framework~\cite{Bellifemine.2001} which was also used in the application of multi-agent systems to HVAC control~\cite{Constantin.2016, Huber.2017}.

Implementing this complex common communication framework requires considerable effort, whereas auctions offer the advantage of being simple to implement as their scope is restricted to allocating goods~\cite{Wooldrige.2002}.
Nevertheless, they still offer a wide range of possible configurations for ensuring that the desired outcome is the goal towards which the agents converge, such as English, Dutch or Vickrey auctions and the Vickrey-Clarke-Groves mechanism among others~\cite{Vlassis.2007, Wooldrige.2002}.
As the problem of fluid system control fundamentally consists of allocating volumes of fluid to various consumers, the approach of designing a market mechanism is promising due to its simplicity.
The approach does not require extensive physical modeling of the fluid system.

To summarize the above, agent-based systems are a promising approach to distributed control and have been applied to various technical systems related to fluid systems as well as fluid systems themselves, e.g., water distribution systems.
Furthermore, a plethora of work has been carried out investigating various approaches to designing multi-agent systems for distributed control out of which three main strands can be identified: (i) distributed model predictive control, (ii) multi-agent reinforcement learning, and (iii) game theory and mechanism design.

Accordingly, the presented work will investigate the application of distributed model predictive control, the multi-agent actor-attention-critic algorithm for multi-agent deep reinforcement learning as well as a market mechanism for distributed fluid system control.
Each of these approaches will be further detailed in Section~\ref{sec:matmet}.

\subsection{Structure of the Paper}

Having outlined the motivation, illustrated the research question and presented the related work, the remainder of the paper is structured in the following way.

Section~\ref{sec:matmet} will first present the model of the fluid system in general terms, as well as the system topology  for the use-case.
It will then present how fluid system components are modelled as agents.
Finally, it will present in detail the methods and design of the implemented control approaches while also discussing the classification of these approaches with regard to transparency, flexibility, observability and modeling effort as well as illustrating the approach to modelling disruptions in the communication network.

Section~\ref{sec:res} will present the results of the study in the following order.
First, addressing research question \textit{(i)}, the overall performance of the three approaches to distributed control is presented.
Next, findings concerning the influence of model information and information exchange are shown as required by research question \textit{(ii)}.
Finally, research question \textit{(iii)} is addressed with results presented for performance of the two distributed systems under disruption.

In section~\ref{sec:disc} the results are discussed further regarding both the concrete research questions and the hypotheses while also reflecting on lessons learned and further research directions.

The paper is completed by a summary and concluding remarks.

\section{Materials and Methods}
\label{sec:matmet}

In the following, the modelling of the fluid system and the investigated application will be presented. Subsequently, it will be explained how the agents for the different components are modelled before introducing the different control approaches of the multi-agent system.

\subsection{Fluid System Model}
\label{sec:fsm}

A fluid system, e.g. a heating circuit, water supply system or industrial cooling system, consists of the following main elements:
(i) the active components pumps and valves and
(ii) passive components such as pipes and generic resistances.
Pumps add hydraulic power to the system, thereby increasing pressure and delivering a volume flow.
Valves reduce the pressure and dissipate hydraulic power, mostly for control purposes. As active components, they are adaptive:
Pumps are able to switch on and off or adjust their rotational speed; valves are able to adjust the pressure loss by changing the valve position.
Pipes, which are passive elements, connect the components and thus determine the topology, whereby pressure losses occur during the transport of the fluid.
Generic resistances, such as heat exchangers, pipe elbows, etc., cause pressure losses depending on the volume flow in the system.
The system behavior, i.e. the volume flow and pressure at different points as well as the power consumption, can be described depending on the components,
topology and control variables (speeds, valve positions) via a non-linear system model.

\subsubsection{Physical-technical description of the system}
\label{sec:phys_mod}
\begin{figure}[h]
\includegraphics[width=13 cm, trim={0 8cm  0 0},clip]{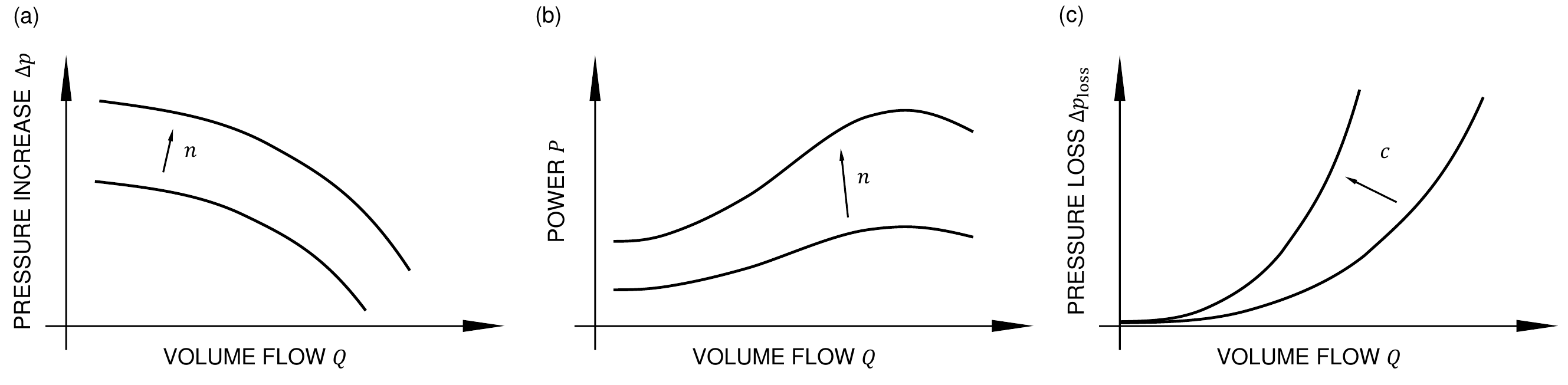}
\caption{Characteristic curves of a centrifugal pump for pressure increase (a) and power demand (b) as well as for a valve (c) \label{fig:component_models}}
\end{figure}

The components are described by their characteristic curves, the machine or system curves.
These are shown qualitatively in Figure~\ref{fig:component_models}.

A pump increases the pressure from the inlet $p_\textrm{in}$ to the outlet $p_\textrm{out}$ by $\Delta p_\textrm{pump}=p_\textrm{out}-p_\textrm{in}$
depending on the volume flow $Q$, speed $n$ and its type (expressed by parameter $\alpha$).
This adds energy to the system. The pressure increase (=head) of a centrifugal pump can be expressed in general by:

\begin{equation}
    \Delta p_\textrm{pump}= \alpha_1 Q^2+ \alpha_2 Q n + \alpha_3 n^2   \label{eq:pump_char}
\end{equation}

The volume flow thus depends on the necessary pressure increase, i.e. the resistance of the system, which causes an interdependence of the components.
The power consumption $P_\textrm{pump}$ is a function of the volume flow $Q$ and the speed $n$:
\begin{equation}
    	P_\textrm{pump}= \beta_1 Q^3+ \beta_2 Q^2 n+ \beta_3 Q n^2  + \beta_4 n^3 + \beta_5 \label{eq:pump_char_power}
\end{equation}
with the parameter $\beta$ depending on the pump type.

If a frequency converter is installed, the speed can be set between minimum and maximum speed or the pump can be switched off: $n \in \{  0\} \cup [n_\textrm{min},n_\textrm{max}]$.

A valve serves as a variable resistance in the system. The pressure reduction (=pressure losses) in a resistance depends quadratically on the flow velocity $w$:
\begin{equation*}
	\Delta p_\textrm{loss}= -\frac{1}{2}  \varrho \zeta w^2=-\frac{1}{2}  \varrho \zeta \left( \frac{Q}{A'}\right)^2=-l Q^2
\end{equation*}
Here, $\varrho$ is the density of the fluid, which in the scope of this study is assumed to be water.
$\zeta$ represents the pressure loss coefficient, which is constant for a component and $A'$ is the cross-section.
The constant terms can be combined to the parameter $l$.
By changing the valve position $v$, the pressure loss coefficient or cross-section can be adapted, causing $l$ and thus the pressure loss to change.
For a valve, $l=l(v) \in [ l_\textrm{min},\infty)$ applies, i.e. there is a certain pressure loss even when the valve is fully open.
When it is fully closed, $l(v)$ becomes arbitrarily large.
The passive components in the system also act as a resistance, but they are not adjustable, i.e. $l_\textrm{passive} = \textrm{const.}$

The system behavior is determined by the components:
The valves and passive components cause pressure losses, which are compensated by the pumps.
The volume flow rates in the system depend on the cumulative resistances.
This creates a complex dependency.
It can be described by the conservation of energy and mass.
In the scope of this study, a model in Dymola \cite{dymola.2020}, which is based on Modelica, is used to simulate the system's behavior.

The control task is to select the pump speeds and valve positions in such a way that all demands
(volume flow rates and pressures at specific points in the system) are satisfied and the energy consumption is as low as possible.
Due to the complexity of fluid systems, heuristics are necessary for the solution, as they are considered in the context of this manuscript.

\subsubsection{Use-Case}

\begin{figure}[h]
\includegraphics[width=18 cm, trim={0 0cm  0 0},clip]{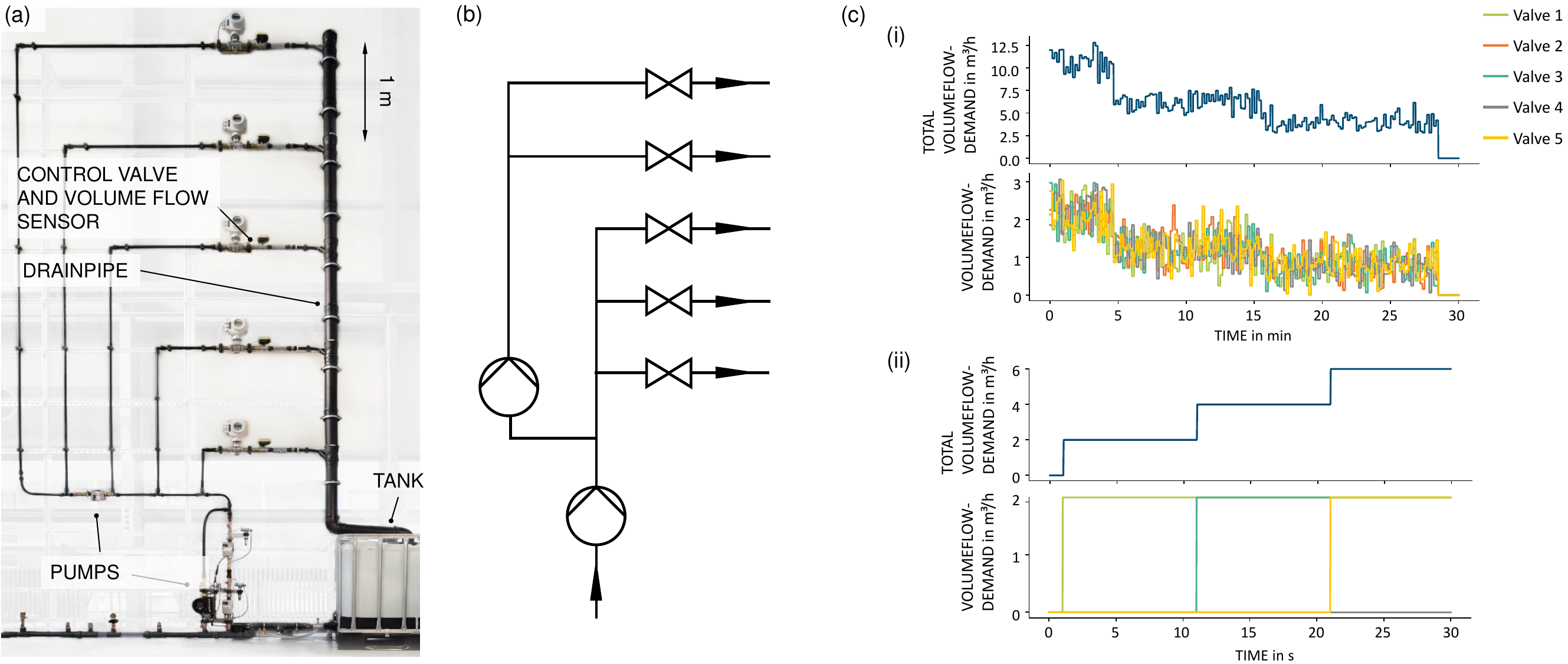}
\caption{Illustration of the use-case with (a) a real-world test rig,
                                           (b) an abstracted model, and
                                           (c) load profiles for (i) \SI{30}{min} and five valves and
                                                                 (ii) \SI{30}{s} and three valves.\textsuperscript{*}}
\scriptsize\raggedright\textsuperscript{*}Numerical data points for the load profiles: \href{https://tudatalib.ulb.tu-darmstadt.de/bitstream/handle/tudatalib/3625.3/figure_3_values.json}{https://tudatalib.ulb.tu-darmstadt.de/bitstream/handle/tudatalib/3625.3/figure\_3\_values.json}
\label{fig:use_case}
\end{figure}

As an application, we consider the water supply of a building, for which a decentralized pressure booster station is used.
For the scaling of the use case, an existing test rig for decentralized Booster stations is used as a reference,
since it is intended to validate the results experimentally in the future, cf. Figure~\ref{fig:use_case} (a).
An in-detail description of the test rig can be found in \cite{Muller.2021b}.
From this, we derive a system with five valves and two pumps, cf.~\ref{fig:use_case}~(b).

Each valve represents a pressure zone, i.e. several floors in the building. The water demand on these floors is summarized in the demand of the valves.
For the evaluation, a 30-minute load profile is used, which simulates the real water demand during 24 hours.
For this purpose, the demand distribution for central booster stations from \cite{Hirschberg.2014} is used and disaggregated into the demand of the different floors.
The resulting water demand for the five valves is shown in Figure~\ref{fig:use_case} (c), (i).
Furthermore, a short load profile of 30 seconds is  used to analyze the behavior of the agents, cf.  Figure~\ref{fig:use_case} (c), (ii).

\subsection{Modelling Fluid System Components as Agents}
The aim of a control system is now to select the control variables in such a way that a common goal is achieved in the best possible way:
to fulfil the volume flow requirements while consuming as little energy as possible.
In distributed control, this is achieved by designing local component controlers as well as rules for their interaction for system-wide control.
This section presents the component level control design, while Section~\ref{sec:mas} will describe the control methods for system-wide control.

The active components of the studied system, pumps and valves are each assigned an agent.
All agents have sensors for observing the environment and actuators for influencing the environment.
The disparate nature of these components requires defining respective types of agents.
The following will present the modeling of each of the agent types used in the presented work.

\subsubsection{Pump Agent}
The pump agent is responsible for the control of a pump.
Depending on the application, the pump can have a controller that sets a target volume flow or be controlled directly by setting a value for relative speed.

The pump can either be switched off, which corresponds to a manipulated variable of zero, or operated in a range between minimum speed $n_{i,\text{min}}$ and maximum speed $n_{i,\text{max}}$.
The manipulated variable range is thus $n_{i} \in \{0\} \cup [n_{i,\text{min}}, n_{i,\text{max}}] = \mathbb{P}$.
In case the pump is controlled directly by the agents, the manipulated variable $u$ corresponds to the pump speed:

\begin{equation*}
    u_{i,t} = n_{i,t}, \quad \forall t \in \mathcal{T}, i \in \mathcal{P},
\end{equation*}

with the set of all time steps $\mathcal{T}$ and the set of all pump agents $\mathcal{P}$.
In case the pump is controlled by a volume flow controller (e.g. a PI-controller) and the agents set the target volume flow, the manipulated variable is

\begin{equation*}
    u_{i,t} = k_{\mathrm{pump}}(Q_{i,t}), \quad \forall t \in \mathcal{T}, i \in \mathcal{P}.
\end{equation*}

Here, $k_{\mathrm{pump}}: \mathbb{R}_{0,+} \rightarrow \mathbb{P}, Q_{i,t} \mapsto u_{i,t}$ is a controller function which maps the required volume flow to a value for the manipulated variable.
Other controllers may be necessary for the motors inside the pump to implement the value of~$u_{i,t}$, but are not considered within the scope of this work.

The observations of pump agents can be volume flow rate as well as electrical power.
The electrical power depends on the speed and the volume flow and usually follows a non-linear characteristic curve (cf.~\ref{sec:phys_mod}).
The deliverable volume flow  also depends on the pressure difference.
Since the aim of the pump is to deliver the required volume flow at minimum electrical power, only the electrical power is mapped with the cost function.
A minimum electrical power $P_{i,\text{min}}$, even if the rotational frequency is zero, comes from features such as a display or standby functionality.
The maximum electrical power $P_{i,\text{max}}$ provides a limit on the pump's performance.
In order for the cost of the agents to be in a similar range, the cost is already related to the maximum power in the cost function.
Weighting against the costs of the pumps against each other or against the valve agents can then be done with the dimensionless weighting factor $\lambda_i$.
Thus, the costs are expressed as

\begin{align}
    c_{\text{pump},i,t} & = \lambda_i \cdot \left(\frac{P_{i,t} - P_{i,\text{min}}}{P_{i,\text{max}}} \right) ^2, \quad \forall t \in \mathcal{T}, i \in \mathcal{P}.
\label{eq:Kosten_Pumpe}
\end{align}

\subsubsection{Valve Agent}

The valve agent is responsible for controlling its valve.
Similar to the pump agents, the valves can have a volume flow controller or be controlled directly by the valve agent setting the valve opening $v$.
The range of the valve opening is $v_{i} \in [v_{i,\text{min}}, v_{i,\text{max}}] = \mathbb{V}$.
In case the agent directly manipulates the valve opening, the relation is

\begin{equation*}
     u_{i,t} = v_{i,t}, \quad \forall t \in \mathcal{T}, i \in \mathcal{V},
\end{equation*}

with the set of all valve agents $\mathcal{V}$.
Analogous to the pump agents, in case the valve is controlled by a volume flow controller (e.g. a PI-controller) and the agents set the target volume flow, the valve opening is

\begin{equation*}
    u_{i,t} = k_{\mathrm{valve}}(Q_{i,t}), \quad \forall t \in \mathcal{T}, i \in \mathcal{V},
\end{equation*}
with $k_{\mathrm{valve}}: \mathbb{R}_{0,+} \rightarrow \mathbb{V}, Q_{i,t} \mapsto u_{i,t}$, the controller function which maps the required volume flow to a value for the manipulated variable.

Relevant variables for the valve agents are the volume flow as well as the differential pressure.
A volume flow or pressure demand is registered directly with the valve agents. \\
Therefore, the cost function consists of the deviation between the target volume flow and the target pressure.
Since only the volume flows are considered in this work, the cost function for the valve agents with the volume flow demand $Q_{i,t,\text{demand}}$ and the actual volume flow $Q_{i,t,\text{actual}}$ results in

\begin{align}
    c_{\text{valve},i,t} & = \lambda_i \cdot (Q_{i,t,\text{demand}} - Q_{i,t,\text{actual}})^2, \quad \forall t \in \mathcal{T}, i \in \mathcal{V} .
\label{eq:Kosten_Ventil}
\end{align}

The cost factor $\lambda_i$ is given for the valve agents $i \in \mathcal{V}$ in units $(\frac{1}{m^3/h})^2$ to make the cost a dimensionless quantity.
Squaring the deviation from the target flow rate is also intended to ensure that low total costs correspond to a fair solution for all valves.
Agents should not choose to disregard a valve whose target fulfilment has a high cost.

\subsection{Control Methods of Multi-Agent System}\label{sec:mas}

Shifting the focus now from the component level to the system level,
the following section will present the three separate control methods implemented, which ensure the agents interact to reach the target of the control task.
It further includes a description of conventional control methods.
Finally, various characteristics will be introduced which will serve to assess the implemented approaches with regard to aims discussed in Section~\ref{sec:intro}.

\subsubsection{Distributed Model Predictive Control}
\label{sec:dmpc-meth}

The method for distributed model predictive control implemented in the presented work follows the work of Pannocchia closely and can be found in detail there~\cite{Pannocchia.2013b}.
The most fundamental equations and basic algorithms are outlined in the following.

Two assumptions are made: \textit{(i)} the system state is equal to the output and \textit{(ii)} the system dynamics are neglected, i.e., the system state does not depend on the preceeding system state.
Applying the assumptions to the non-linear fluid system described above yields the non-linear steady state system

\begin{equation*}
    \mathbf{y} = g(\mathbf{u}),
\end{equation*}

with the steady state model
$g: \mathbb{V}^{\lvert \mathcal{V} \rvert} \times \mathbb{P}^{\lvert \mathcal{P} \rvert}
    \rightarrow \mathbb{R}_{0+}^{{\lvert \mathcal{V} \rvert} \times {\lvert \mathcal{P} \rvert}},
    \mathbf{u} \mapsto \mathbf{y}.$
Here, $\lvert \mathcal{V} \rvert, \lvert \mathcal{P} \rvert$ describe the cardinality of the set of valve agents $\mathcal{V}$ and pump agents $\mathcal{P}$, respectively.
The vector $\mathbf{u} = [u_1, u_2, ..., u_A]^T$, with $A = \lvert \mathcal{A} \rvert$ the number of agents, contains the manipulated varibles of all agents $\mathcal{A} = \mathcal{V} \cup \mathcal{P}$.
The vector $\mathbf{y} = [y_1, y_2, ..., y_A]^T$ contains the output of all agents, i.e. values of volume flow and power for valve and pump agents respectively.
For attaining $g$, the substitute model of the fluid system, a neural network of the type multi-layer perceptron with a sigmoid function was programmed using the python library \emph{scikit-learn}.
Using the Dymola model of the use-case test rig, randomized combinations of input data with $u_{i,\mathrm{min}} \leq u_i \leq u_{i,\mathrm{max}}$ and the corresponding outputs were generated as a set of training data for the neural network, where 80\% served as training data and 20\% as validation data.

Model predictive control requires consideration of system dynamics which contradicts the second assumption made above.
However, by introducing a restriction in the values the manipulated variable of each agent can take in each time step $u_{i,t}-\delta \leq u_{i,t+1} \leq u_{i,t}+\delta$,
the concept of prediction can be introduced into the steady state model.
This concept is detailed in the following.
The vector of manipulated variables of all agents $i \in \mathcal{A}$ at time $t \in \mathcal{T}$, the set of all time steps, can be written as

\begin{equation*}
\mathbf{u}_t  = [\mathbf{u}_{i,t}, \mathbf{u}_{-i,t}]^T,\quad i, -i \in \mathcal{A}, t \in \mathcal{T},
\end{equation*}

where $i$ denotes Agent $i$ and $-i$ denotes all agents apart from Agent $i$.
The manipulated variable vector for each agent consists of $L$ entries with $L$ as the number of predicted control steps (also known as prediction horizon), such that

\begin{equation*}
\mathbf{u}_{i,t} = [u_{i,t} (0), u_{i,t} (1), ..., u_{i,t} (L-1)]^T,\quad i \in \mathcal{A}, t \in \mathcal{T}.
\end{equation*}

Only the first value is implemented, though, before a new control step calculates $\mathbf{u}_{i,t}$ afresh with a receding horizon.
The substitute model can then be written as

\begin{equation*}
\mathbf{\hat{y}}_{t+1} = g([\mathbf{u}_{i,t}, \mathbf{u}_{-i,t}]),
\end{equation*}
with the predicted output of the next time step $\mathbf{\hat{y}}_{t+1}$.

Each agent only considers $\mathbf{u}_{i,t}$ as a variable, while $\mathbf{u}_{-i,t}$ are considered constants.
However, during calculation of each control step, several iterations can be carried out where the agents exchange information about their proposed value of $\mathbf{u}_{i,t}$.
These communication rounds help to improve the overall solution, yet also increase computational effort.
All agents have a common objective, with the overall cost function calculated as

\begin{equation*}
    V(\mathbf{\hat{y}}_{t+1}) = \sum_{i=1}^A\ \rho_i \hat{C}_i (\mathbf{\hat{y}}_{i,t+1}),
\end{equation*}

with a weighting factor $\rho_i$, where the cost function of each agent is calculated for the entire prediction horizon, such that

\begin{equation*}
    \hat{C}_i = \sum_{j=1}^L\ c_i(\mathbf{\hat{y}}_{i,t+1}(j))
\end{equation*}

with $c_i$ being the cost function of the pump and valve agents as defined in Eq.~\eqref{eq:Kosten_Pumpe} and Eq.~\eqref{eq:Kosten_Ventil} respectively.
The overall optimization problem for each agent can then be written as

\begin{mini*}
{\mathbf{u}_{i,t}}{V(\mathbf{\hat{y}}_{t+1})=\sum_{i=1}^A\ \rho_i \hat{C}_i (\mathbf{\hat{y}}_{i,t+1})}{}{}
\addConstraint {\mathbf{\hat{y}}_{t+1}}
               {=g([\mathbf{u}_{i,t}, \mathbf{u}_{-i,t}]),\quad}
               {i, -i \in \mathcal{A}, t \in \mathcal{T}.}
\end{mini*}

In the presented work, a central model predictive control is also implemented as a benchmark for the distributed model predictive control.
The method follows that outlined above with the difference that there is just one agent optimizing the control variables for all components with equal weights of $\rho_i = 1$.
The optimization problem is thus given as

\begin{mini*}
    {\mathbf{u}_{t}}{V(\mathbf{\hat{y}}_{t+1})=\sum_{i=1}^A\ \hat{C}_i (\mathbf{\hat{y}}_{i,t+1})}{}{}
    \addConstraint {\mathbf{\hat{y}}_{t+1}}
                   {=g(\mathbf{u}_{t}),\quad}
                   {t \in \mathcal{T}.}
\end{mini*}

\subsubsection{Multi-Agent Reinforcement Learning}
The Reinforcement Learning problem is written as a Markov game, consisting of action, states and rewards.
The optimization problem of each agent $i \in \mathcal{A}$ for each time $t \in \mathcal{T}$ is to find an action that minimizes the costs

\begin{mini*}
    {\mathbf{u}_{i,t}}{\sum_{j=1}^L c_{t+j}}{}{}
    \addConstraint {\text{Eq.~\eqref{eq:actions_discrete}}.}
                   {}
                   {}
\end{mini*}

\textit{States:}
Agent $i$ chooses its action based on observations $o_i \in \mathcal{O}$, which represents a subset of all states $s \in \mathcal{S}$.
In addition to the states, the agents also observe the volume flow requirements during training.
An agent observes volumeflow requirements of valve agents of the set $\mathcal{V}$.
At time $t$, all observations are $o_t = \{o_{1,t}, o_{2,t},...,o_{A,t}\}$.

Two versions of MADRL are used in this work. In centralized MADRL, each agent $i \in \mathcal{A}$ has the complete system knowledge. This means that the observed states contain the output variables of all agents, as well as the volume flow requirements of all valve agents:
\begin{equation*}
o_{i,t} = \{y_{z,t}|_{\forall z\in\mathcal{A}}, Q_{z,t,\text{user}}|_{\forall z\in \mathcal{V}}\}\;.
\end{equation*}

In decentralized MADRL, the agents have only limited system knowledge. The observed states are only their own output variables and the volume flow requirements of all valve agents:
\begin{equation*}
o_{i,t} = \{y_{i,t}, Q_{z,t,\text{user}}|_{\forall z\in \mathcal{V}}\}\;.
\end{equation*}

\textit{Actions:}
The pump speeds and valve positions are discretized to create a discrete action space, the manipulated variables of an agent $i$ thus become
\begin{equation}
    u_{i,t} \in \{u_i^1, u_i^2, ..., u_i^M\}, \quad \forall t \in \mathcal{T},\quad i \in \mathcal{A} \; .
    \label{eq:actions_discrete}
\end{equation}
The number of discrete valve positions or rotational speed is the same for all agents and is denoted by $M$.
The actions of all agents are the control variables $a_t = \{u_{1,t}, u_{2,t}, ..., u_{A,t}\}$.

\textit{Rewards:}
Since training is centralized in MAAC and the rewards are also only necessary for training, the global reward is used for all agents, i.e. $r_{i,t} = r_{1,t} = r_{2,t} = ... = r_t$.
To compare the different control methods, the global rewards are formed from the global costs with $r_t = -c_{\text{global},t}$.
All of this provides the training tuple $\langle s_{t}, a_t, s_{t+1}, r_{t+1} \rangle$.

MAAC \cite{Iqbal.2018} is an actor-critic method where both policy and value functions are learned.
The policy function calculates the action (Actor) and the Value function evaluates that action (Critic).
As training progresses, Actor and Critic both improve.

The value function is the action value function $\chi(s_t,a_t)$, which is approximated by a neural network $\chi_{\psi}(s_t,a_t)$, where the neural network is parameterized by the weights $\psi$.
The policy function is also approximated by a neural network $\pi_{\theta}(a|s)$ with weights $\theta$.

To stabilize the training, in addition to the neural network just mentioned, so-called target networks $\chi_{\overline{\psi}}(s_t,a_t)$ and $\pi_{\overline{\theta}}(a|s)$ are used, whose weights are updated during the training.
$\mathcal{D}$ denotes the experience replay buffer, where all previous experiences (i.e. tuples of states, actions, new states, and rewards) are stored.
To make training more efficient, each agent utilizes an attention mechanism, where attention weights are computed that indicate the contribution of the other agent's observations to the action value function.
With the attention mechanism, it is possible for the agent to dynamically focus on other agents during training.

During the centralized training phase, the critics neural networks minimize the following loss function \cite{Yu.2021}:
\begin{align*}
\begin{split}
\mathcal{L}_{\chi}(\psi) &= \sum_{i=1}^{A} \mathbb{E}_{(o, a, o', r) \sim \mathcal{D}}\left[\left(\chi_{i}^{\psi}(s, a)-\xi_{i}\right)^{2}\right], \\
\xi_{i} &= r_{i}(o, a)+\gamma \mathbb{E}_{a' \in \pi_{\bar{\theta}}(o')}[\chi_{i}^{\bar{\psi}}(o', a')-\varphi \log(\pi_{\bar{\theta}_{i}}(a'_{i},o'_{i}))]
\end{split}
\end{align*}

$\gamma$ is the discount factor and $\varphi$ is the temperature parameter that can be used to adjust the exploration during training. \\
For the policy function, a policy gradient method is used with the gradient \cite{Yu.2021}
\begin{align*}
\begin{split}
\nabla_{\theta_{i}} J(\theta)&=\mathbb{E}_{o \sim \mathcal{D}, a \sim \pi}\left[\nabla_{\theta_{i}} \log \left(\pi_{\theta_{i}}\left(a_{i} \mid o_{i}\right)\right) \varepsilon_{i}\left(o_{i}, a_{i}\right)\right], \\
\varepsilon_{i}\left(o_{i}, a_{i}\right)&=-\varphi \log \left(\pi_{\theta_{i}}\left(a_{i} \mid o_{i}\right)\right) + A_{i}(o, a)
\end{split}
\end{align*}

where $A_{i}(o, a)$ is the advantage-function, further explained in \cite{Yu.2021}.

In the execution phase, the actions are calculated with the learned policy function from the observations.
For a more detailed description of the implementation, the reader may refer to \cite{Yu.2021} and \cite{Iqbal.2018}.

\subsubsection{Market Mechanism}
In order to achieve the overall system-wide goal while ensuring autonomous decision-making for the agents, a market mechanism is developed as a third approach.
A set of market and agent rules are defined which agents have to follow.
Within these rules, a market with supply and demand is established.

The agents buy and sell guarantees for a certain volume flow.
In turn, they must pay for energy costs or are paid to provide a service.
For example, after a sale, pump-agents guarantee that they will pump a certain volume flow.
To do this, they must estimate how "expensive" the necessary pressure increase will be and, if necessary, also purchase flow guarantees from upstream components.
The agents act according to the market rules and aim to maximize their own profit.
Thus, energy-efficient behavior is encouraged. This corresponds to the basic idea of a free market, which is able to control the complexity of the system even without a perfect system model and central control.

\noindent The market rules are:
\begin{enumerate}
    \item volume flow guarantees are the subject of selling/buying
    \item each agent must deliver the sold volume flow at its output
    \item each agent can only interact with its neighboring agents
    \item each agent has a finite budget available per time step, which consists of the costs incurred by not achieving the goal
    \item the order of buying is from down- to upstream, starting with the lowest agent
\end{enumerate}

In the method, negotiation rounds are triggered regularly. In these rounds, the agents first calculate their demands and offers and second purchase and sell volume flow guarantees. In between the negotiation rounds, there is a control phase in which the agents control the volume flow to the sum of purchased and sold guarantees.

The procedure of a negotiation round is shown in Algorithm \ref{algo:negotiation}.
In each  round, the individual agents buy a  volume flow guarantee, for which they request offers from all upstream neighboring agents and buy the best one.

\begin{algorithm}[h]

    \SetAlgoLined
	\SetKwInOut{Input}{input}
	\SetKwInOut{Output}{output}

	\Input{~Volume flow demand $Q_{\textrm{demand}, i}$ of all agents $i\in \mathcal{A}$}

	\Output{~Volume flow set points $Q_{\textrm{set}, i}$ for the local controller  of all agents $i\in \mathcal{A}$}

	\vspace{0.3em}
	$Q_{\textrm{set}, i} \leftarrow 0 ~ \forall i \in \mathcal{A}$

	\vspace{0.3em}

        \For(\tcp*[f]{ordered from upstream to downstream agents}){$i=1$ \KwTo $|\mathcal{A}|$ \
        }{

        \vspace{0.3em}

            \If(\tcp*[f]{procedure at agent level}){$Q_{\mathrm{demand}, i} > 0$}{
            calculate its budget $B_i$

            \vspace{0.3em}

             ask for offers $O \leftarrow O_{i_\textrm{up}}$ from all upstream agents $i_\textrm{up} \in \mathcal{A}_\textrm{up}$

             \vspace{0.3em}

             choose best offer: $(Q_{\textrm{purchased},i}, P_{\textrm{purchased},i}) = \textrm{best}(O)$

             accept and pay for accepted offer: $B_i =  B_i -  P_{\textrm{purchased},i}$

			calculate new demand and setpoint:

			$Q_{\textrm{demand}, i} = Q_{\textrm{demand}, i} - Q_{\textrm{purchased},i}$

			$Q_{\textrm{set}, i} =  Q_{\textrm{purchased},i}$

            }
    }

\caption{Procedure of a negotiation round triggered at time $t$.}\label{algo:negotiation}
\end{algorithm}

Besides the market rules, the behavior of the agents is  crucial. The goal for each agent is to maximize its profit. For this purpose, the best possible offers have to be calculated and favorable offers have to be purchased.

While the selection of the best offer is trivial, since simply the cheapest offer that meets the demand is selected, the calculation of the offers is more complex.
Since only upstream agents are asked for offers and only pump agents are upstream agents in the use-case model, the method for calculating offers is only implemented for pump agents.
When a pump agent is asked for an offer, it calculates offers for different volume flows according to Algorithm \ref{algo:offers}.
On the one hand, the individual costs $b_{\textrm{ind}}$ and, on the other hand, the costs that arise because guarantees have to be purchased from upstream components $b_{\textrm{purchased}}$ are taken into account.
The individual costs of the pumps arise from the energy required to increase the pressure.
The agent estimates the costs for the total guaranteed volume flow.
Since the agent sells parts of the total volume flow several times to one or more agents, the buying agent is only charged for its portion of the total volume flow.
This means that for the offers, the previous earnings $b_{\textrm{earnings}}$ have to be subtracted from the costs of the total volume flow.

\begin{algorithm}[h]

    \SetAlgoLined
	\SetKwInOut{Input}{input}
	\SetKwInOut{Output}{output}

	\Input{~current operating point ($Q,\Delta p)$, ~previous earnings $b_{\textrm{earnings}}$}

	\Output{~set of offers $O=\{(Q_{\textrm{offer},1}, b_{\textrm{offer},1}),  ..., (Q_{\textrm{offer},N_\textrm{offers}}, b_{\textrm{offer},N_\textrm{offers}}) \}$}

	\vspace{0.3em}

    \For{$j=1$ \KwTo $N_\mathrm{offers}$ \
        }{
        $Q_{\textrm{offer},j} = j \cdot Q_{\textrm{max}} / N_\textrm{offers} $

        \vspace{0.3em}

        estimate $b_{\textrm{ind},j}(Q_{\textrm{offer},j}, Q, \Delta p)$ according to Eq.~\eqref{eq:cost_pumps_1} - Eq.~\eqref{eq:cost_pumps_last}\\
        \vspace{0.3em}
        ask for offers $O \leftarrow O_{i_\textrm{up}}$ from all upstream agents $i_\textrm{up} \in \mathcal{A}_\textrm{up}$

        choose best offer: $(Q_{\textrm{purchased},j}, b_{\textrm{purchased},j}) = \textrm{best}(O)$

        $ b_{\textrm{offer},j} = b_{\textrm{ind},j} + b_{\textrm{purchased},j} - b_{\textrm{earnings}}$

        }

\caption{Calculation of several offers if a pump agent $i\in \mathcal{P}$ is asked for an offer.}\label{algo:offers}
\end{algorithm}

The estimation of the individual pumping costs is crucial for achieving efficient system operation.
In this work, it is realized by rule-based estimations of the presumed power demand for the delivery of a certain volume flow.
Therefore, the pressure increase and necessary speed are estimated. For the system resistance, a quadratic relationship is assumed, as the pressure loss depends quadratically on the volume flow. The pump agent $i \in \mathcal{P}$ estimates the system characteristics for the current  operation point based on the measured volume flow $Q_{i}$ and pressure increase $\Delta p_{i}$:
\begin{equation} \label{eq:cost_pumps_1}
    l_{\textrm{est}, i}= \Delta p_{i} / Q_{i}^2;
\end{equation}

From this, the necessary pressure increase $\Delta p_{\textrm{est},i}$ is calculated for the offered volume flow $Q_{\textrm{offer},i}$:

\begin{equation*}
   \Delta p_{\textrm{est},i} =  l_{\textrm{est}, i} Q_{\textrm{offer},i}^2 ;
\end{equation*}

This is equal to the pressure that the pump has to provide according to the characteristic curve, cf. Eq.~\eqref{eq:pump_char}:
\begin{equation*}
    l_{\textrm{est}, i} Q_{\textrm{offer}, i}^2 =  \alpha_{\textrm{1}, i} Q_{\textrm{offer}, i}^2+ \alpha_{\textrm{2}, i} Q_{\textrm{offer}, i} n_{\textrm{est}, i} + \alpha_{\textrm{3}, i} n_{\textrm{est}, i}^2  ;
\end{equation*}
which can be solved for the estimated speed $n_{\textrm{est}, i}$.

Then, using the estimated speed and the power curve, cf. Eq.~\eqref{eq:pump_char_power}, the required power $P_{\textrm{est}, i}$ is estimated:
\begin{equation} \label{eq:cost_pumps_last}
   P_{\textrm{est}, i} =  \beta_{\textrm{1}, i} Q_{\textrm{offer}, i}^3+ \beta_{\textrm{2}, i} Q_{\textrm{offer}, i}^2 n_{\textrm{est}, i} + \beta_{\textrm{3}, i} Q_{\textrm{offer}, i} n_{\textrm{est}, i}^2 + \beta_{\textrm{4}, i} n_{\textrm{est}, i}^3 +\beta_{\textrm{5}} ;
\end{equation}

By multiplying with the constant $d_\textrm{power}$, this is converted into dimensionless individual costs: $b_{\textrm{ind}, i} = d_\textrm{power} \cdot P_{\textrm{est}, i}$.

In order to guarantee the purchased or sold volume flows after the negotiation, a controller is required. For this purpose, a conventional PI controller is used. 
For the pumps, a special feature is introduced, which is called artificial scarcity: A slightly lower volume flow than sold is set. This ensures that the downstream components behave "efficiently". In the case of the valves, they will not generate any unnecessary pressure loss, since they want the highest possible volume flow. This prevents that e.g. all valves have only a very low opening and the pumps have to work against it.

In this approach, communication is only necessary during negotiation rounds and only between direct neighbors. A system-wide model is not required, but an estimation of the system characteristics is. A wrong estimation can lead to a bad calculation of the offers and thus low efficiency, but the functionality is not affected.

\subsubsection{Conventional Methods}
There is typically no communication when conventional methods are used to control the pumps and valves, i.e. there is local control according to Figure~\ref{fig:concepts}.
In most cases, the valves are responsible for controlling the actual process function, e.g. volume flow or temperature.
The pumps must ensure that sufficient pressure is supplied at the valves so that they can operate safely.
To do this, the pumps anticipate the operation of the valve by controlling the pressure at the pump outlet as a function of the volume flow.
The pressure must be high enough to compensate for all pressure losses in the system all the way to the valves, for which a so-called pressure control curve is used.
This always results in overfulfilment, which leads to energetic losses.

In the context of this work, the valves locally control a flow rate  using a PID controller independently of each other.
For pumps, the following control approaches are considered for comparison purposes:

\begin{itemize}
    \item non-controlled: all pumps run continuously at their maximum speed
    \item constant pressure control: the pressure difference of the pumps is constant $\Delta p = \Delta p^\text{set}_\text{const}$
    \item Proportional pressure control: the pressure difference of the pump is proportional to the volume flow, in addition to a constant part $\Delta p = l_\text{prop} Q + \Delta p^\text{set}_\text{const}$
\end{itemize}

The respective values of the parameters are estimated using a model of the fluid system. For this purpose, a design flow rate $Q^\text{design}$ must be assumed, which can be covered maximally by the system. If a higher value $Q^\text{design}$ is assumed than the expected maximum volume flow $Q^\text{max}$, this corresponds to the integration of a safety factor $S$: $Q^\text{design} = S \cdot Q^\text{max}$.

\subsubsection{Assessment Criteria}

This section will serve the purpose of presenting assessment criteria for the control methods based on multi-agent systems.
It will further discuss various characteristics of the approaches, helping to categorize them.

The assessment criteria are derived from the challenges presented in Section~\ref{sec:intro}:
\begin{itemize}
 \item[(i)] fulfilling the functionality of the system, i.e., the control objective of providing a given volume flow to valves,
 \item[(ii)] effort involved for fulfilling the functionality, i.e., consumption of electrical energy during operation as well as modelling effort of generating a substitute model (DMPC), training of agents (MADRL), or designing a mechanism (market mechanism) and information exchange between agents,
 \item[(iii)] availability, i.e., reliability of fulfilling the functionality in case of disruption, and
 \item[(iv)] acceptability, i.e., transparency of control decisions as well as portability, flexibility, and scalability of the approaches.
\end{itemize}

The cost functions for valves and pumps yield numerical values for assessing fulfillment of functionality and effort with regard to energy consumption (cf. Section~\ref{sec:res}).

Modeling effort, by contrast, is harder to gauge.
DMPC requires a substitute model, which in the presented work is acquired using machine learning methods.
The implementation of MADRL in this study uses model-free rather than model-based approaches, where the system model is learned directly by the agents.
Nevertheless, a new substitute model needs to be learned for every new application or change in the system configuration.
Market mechanism is inherently model-free, though the pumps do require a model for calculating the cost of providing different volume flows and the market rules need to be designed and tested prior to implementation.

Regarding information exchange, observability is a category with which multi-agent systems can be classified.
According to Vlassis, full observability is given when each agent perceives the entirety of states that constitute the state of the agents' environment with each observation~\cite{Vlassis.2007}.
Partial observability is given when the agents only receive partial information on the current state of their environment with each observation~\cite{Vlassis.2007}.
This study considered both fully observable, termed centralized, and partially observable MADRL.
As stated above, in DMPC there is information exchange between agents on their proposed values for the manipulated variable, with several iterations being possible.
In market mechanism, the information exchange is limited to offers and purchases of volume flow and restricted to neighboring agents.
Thus, effort of information exchange strongly depends on the control approach.
What effect the effort of information exchange has on the fulfillment of the function is the subject of research question (ii).

For assessing the availability, the same cost functions are used as for assessing fulfillment of functionality.
The values of these cost functions for a disrupted system are compared to those of an undisrupted system.
How the disruption is modelled in this study is presented in the subsequent section.

For gauging acceptability, the approaches can be arranged on the scales of black box vs. white box and flexibility.
In market mechanism, all control decisions and market rules are transparent and easily understandable, i.e. this represents a white box approach.
The machine learning methods of MADRL, by contrast, don't allow for control decisions to be comprehended or traced and neither is the learned substitute model discernible, which makes this a black box approach.
DMPC is comprehensible in the regard of the employed optimization methods, though the substitute model created using machine learning is not transparent, meaning that this approach is located between white box and black box.
Market mechanism is also the most flexible design approach, as it allows agents to be added or removed without changing the system model as long as the communication between upstream and downstream neighbors is ensured.
Both DMPC and MADRL depend on the learned substitute model which requires new training rounds once the system is adapted, rendering the approaches less flexible.

To summarize, the requirement of a substitute model plays an important role when assessing both effort and acceptability associated with control approaches.
Observability has further implications for effort.
Transparency of control decisions and flexibility of control approaches with regard to adapting systems are important criteria for assessing acceptability.
Cost functions of agents enable the assessment of functional performance and availability.
For the latter, modeling disruptions plays an important role which is presented in the following.

\subsection{Modelling Disruptions}
\label{sec:disrmod}

A prerequisite for transitioning from local control of fluid system components to central or distributed control architectures is the implementation of information and communication infrastructure.
This allows information to be passed from components to a central control unit or between individual component agents in the case of distributed control.
It has been remarked that increased dependence on communication infrastructure of fluid systems and water distribution systems in particular introduces new vulnerabilities~\cite{Zimmerman.2020}.
In consequence, the approaches to distributed control using DMPC and MADRL are investigated with regard to their performance faced with a disruption of the communication between agents.
This allows to address the hypothesis regarding reliability of distributed versus centralized control and contributes to answering research question (iii), cf.~\ref{sec:rq}.

\begin{figure}[h]
\center
\includegraphics[width=15 cm]{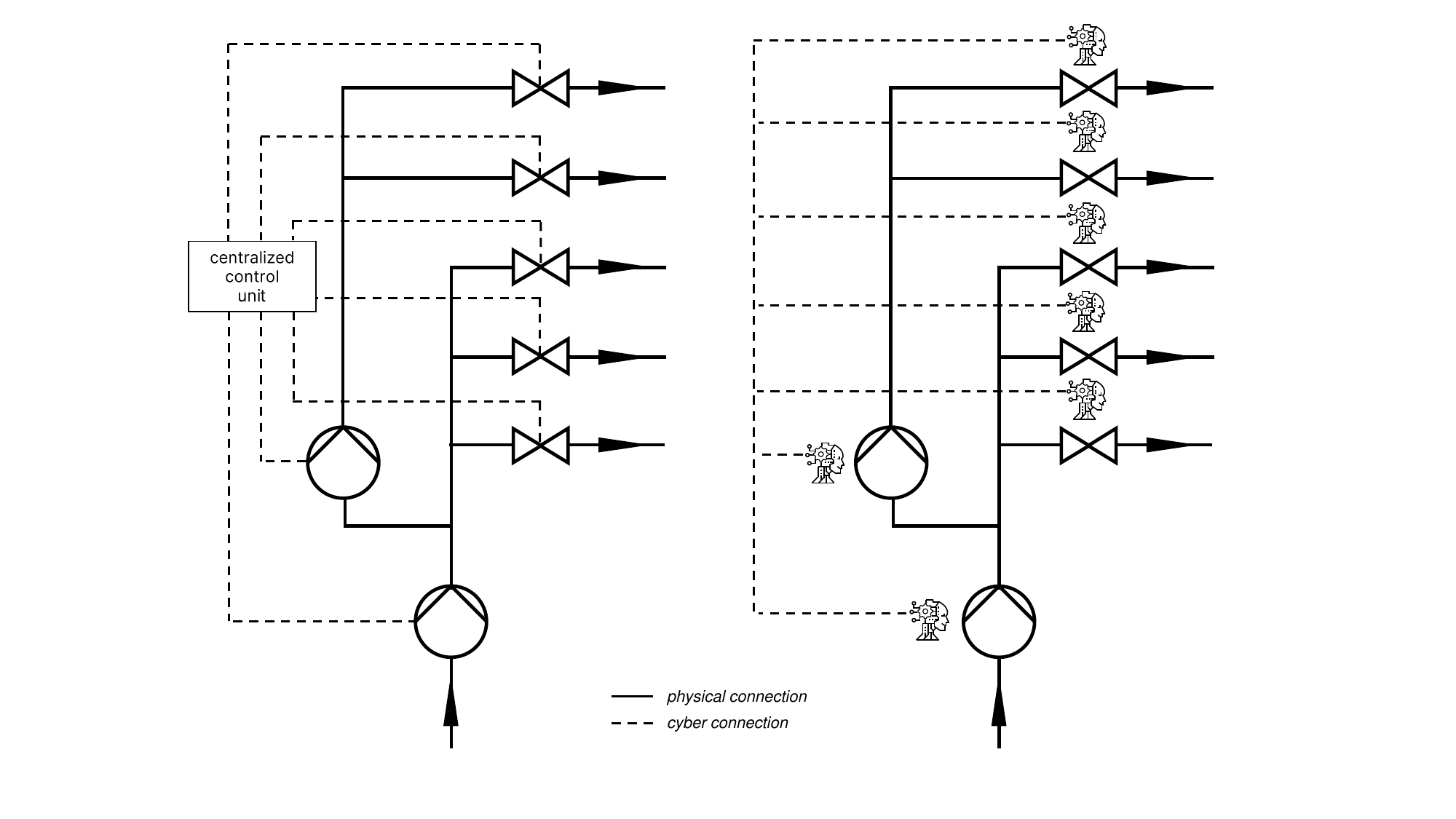} 
\caption{Centralized and distributed control architectures with undisrupted communication infrastructure.}
\label{fig:undisrupted}
\end{figure}

Figure~\ref{fig:undisrupted} shows the communication infrastructure of both the centralized and the distributed control architectures in the undisrupted case.
For centralized control as in Figure~\ref{fig:concepts} (b), the state variables of the components are communicated to the central control unit which in turn communicates the calculated values for the manipulated variables to the components where they are implemented.
For distributed control as in Figure~\ref{fig:concepts} (c), the agents calculate and implement their control variables on a local level.
They further communicate their state variables to one another.
This requires interconnection of communication between all agents.

\begin{figure}[h]
\center
\includegraphics[width=15 cm]{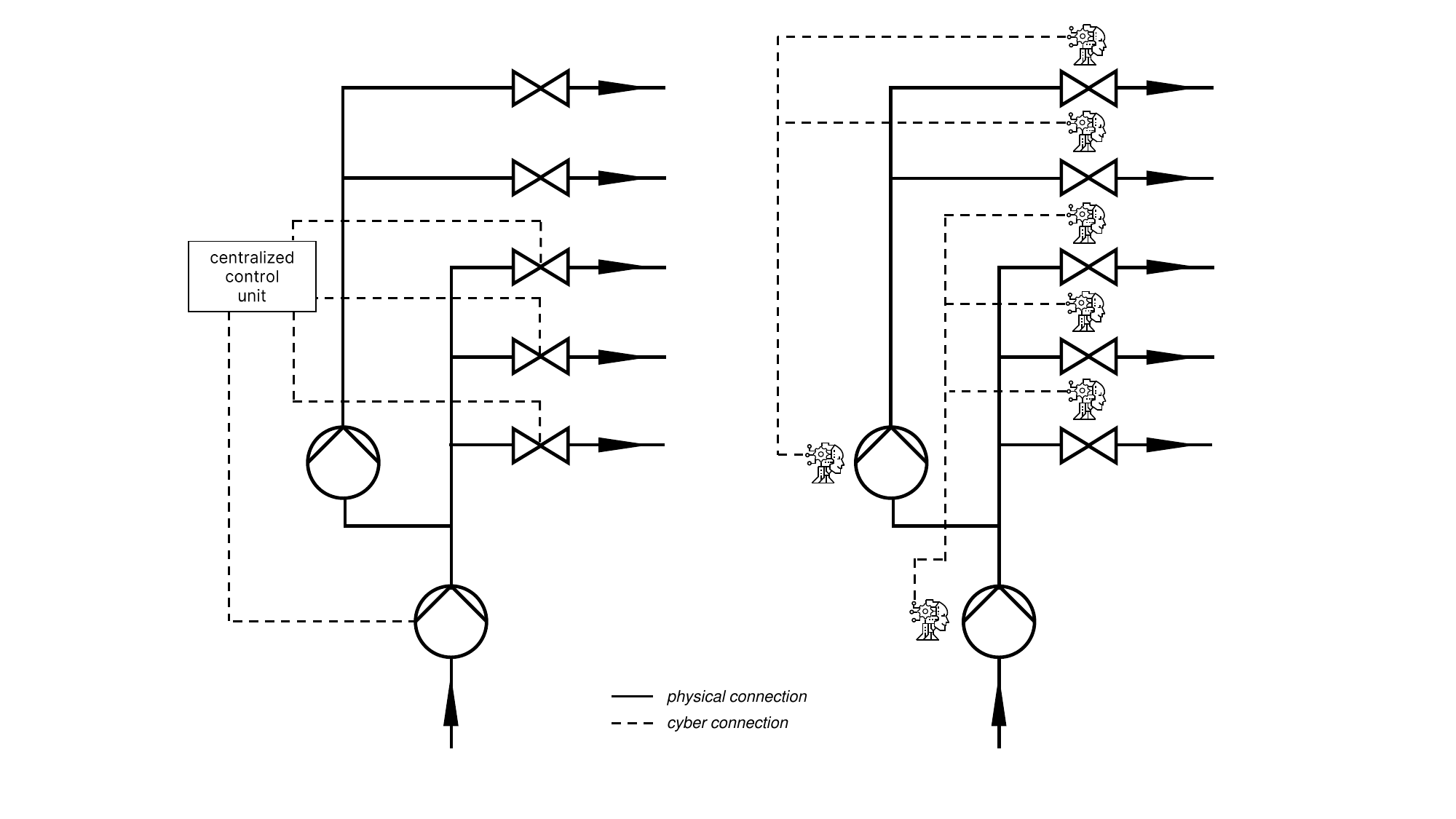} 
\caption{Centralized and distributed control architectures with disrupted communication infrastructure.}
\label{fig:disrupted}
\end{figure}

Figure~\ref{fig:disrupted} illustrates the implementation of a disruption in the communication infrastructure for both the centralized and distributed control architectures.
The centralized control unit loses its connection with the upper pump and two valves and in consequence ceases to consider them when gathering state variables or calculating and implementing control variables.
For the distributed control, on the other hand, the result of the disruption is a separation into two subsystems.
The agents of each subsystem are able to exchange information on state and control variables amongst themselves yet not across the system boundary.
The lower subsystem consists of one pump and three valves.
It is situated upstream from the second subsystem, consisting of one pump and two valves.
However, the lower subsystem is situated at a lower geodetic height than the second subsystem and is thus in a structurally advantaged position.

The disruption scenario is chosen as a result of domain specific considerations.
The valves represent consumers with volume flow demands.
To satisfy these, a fluid system requires a pressure source, represented by the pumps.
Thus, the link chosen to be disrupted connects two units, each of which is a functional fluid system.
Accordingly, the disruption scenario follows the concept of a system consisting of viable units comparable to a honeycomb structure where each cell is a stable structure in itself regardless of the surrounding system as proposed for resilient systems by Heinimann~\cite{Heinimann.2003}.
With regard to the hypothesis and research question under investigation, the chosen disruption scenario is the most compelling for providing insights.

\subsection{Implementation}
All experiments are performed and analyzed using Python.
For the MADRL, the freely available code from MAAC \cite{Iqbal.2018} was used. To use the hydraulic simulation, the interface OpenAI Gym \cite{1606.01540}, which is widely used in the reinforcement learning field, was used.

The hydraulic simulation model based on a real-world test rig was implemented in Dymola, a simulation and modeling software based on the Modelica modeling language.
The final model was exported as a functional mockup unit (FMU) and used with Python via functional mockup interface (FMI) using the Python library \emph{FMPy}.
Both the Dymola model of the test rig used for this study and the FMU file can be found in the data repository\footnote{\href{https://tudatalib.ulb.tu-darmstadt.de/bitstream/handle/tudatalib/3625.3/Fluid_Model.zip}{https://tudatalib.ulb.tu-darmstadt.de/bitstream/handle/tudatalib/3625.3/Fluid\_Model.zip}}.
The differential equations mapped using Dymola are solved using the Dymola solver dassl, a solver with step size control, where the step size was set to \SI{0.025}{\s}.
The experiments were run on a computer with an IntelCore™ i7-8700 CPU with 3.2 GHz clock speed and 16\,GB RAM.
For all simulations, the weighting parameters of the cost of the valves $i \in \mathcal{V}$ are set to $\lambda_i = 1$ and those of the pumps $i \in \mathcal{P}$ are set to $\lambda_i = 2$.

For the model of (D)MPC, a dataset with \num{150000} data points was sampled from the simulation with randomized inputs. Together with the measured outputs, a deep neural network can be trained to act as the system model during optimization.
The R²-value of the system model is \num{0.9873}, meaning that the model is quite close to the Dymola model.
For optimization, the library \texttt{scipy} with the optimizer \textit{SLSQP} is used.
To avoid a solution that is the local and not the global minimum, in each step the optimization is repeated with random starting values.

The hyperparameters for the MADRL algorithm were manually tuned after several experiments. As will be mentioned later, systematic optimization of hyperparameters offers the greatest potential for further improvement.
As in the original implementation of MAAC, the neural networks for both actor and critic consist of two layers with 256 neurons each.
Each training episode has a duration of \SI{4}{\s} and consists of randomized volume flow demand held for \SI{0.5}{\s}.
This allows a wide range of combinations to be covered during training.
In the training, the simulation frequently crashed due to numerical errors caused by the initially very arbitrary actions of the agents. To prevent this, an additional sparse reward of \num{-25} was added to the global reward for each simulation crash.
Training was continued until no improvement of the reward was seen.
This was the case after approximately 4,000 episodes, which corresponds to \SI{4.45}{\hour} of real-life training.

Just like the other methods, the algorithms required for the market mechanism approach are implemented entirely in Python.

\section{Results}
\label{sec:res}

In the following, the results of the different approaches applied to the example system, shown in Figure~\ref{fig:use_case}, are presented.
First, a comparison of the three approaches is made with regard to functionality and energy efficiency for the \SI{30}{min} load profile.
In addition, the behavior of agents and transparency of control decisions are analyzed using the \SI{30}{s} load profile.
Next, the influence of the models on which the approaches are based and the amount of information exchanged between the agents is examined.
Finally, the behavior in case of disruptions is considered.

\subsection{Comparison of Approaches to Distributed Control}
\label{sec:comp}

\begin{figure}[h]
\center
\includegraphics[width=13 cm, trim={0 9cm 8cm 0cm},clip]{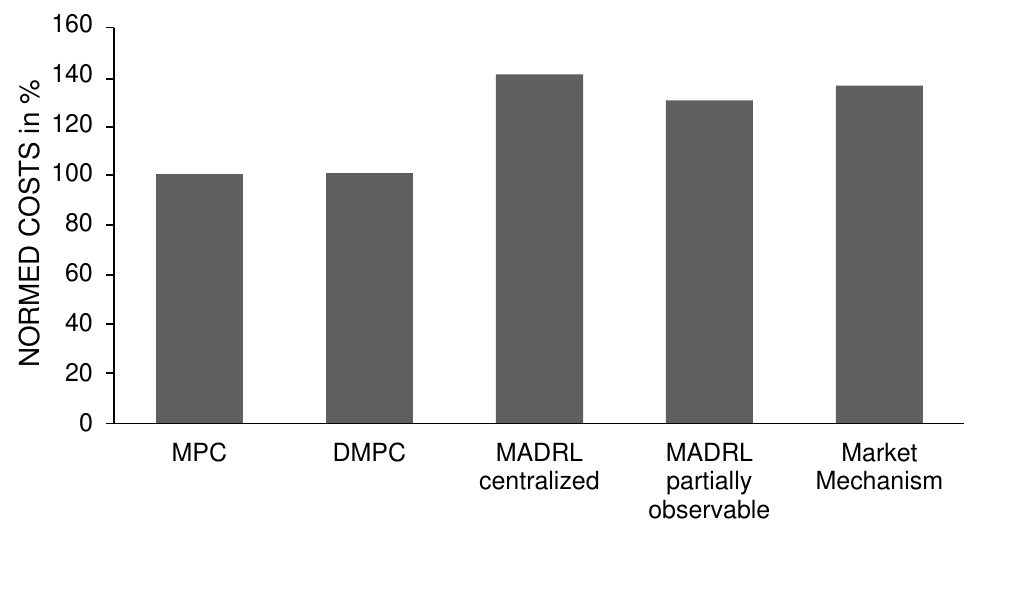} 
\caption{Comparison of costs of Model Predictive Control, Multi-Agent Deep Reinforcement Learning, and Market Mechanism  for the 30 minutes scenario.\textsuperscript{*}}
\scriptsize\raggedright\textsuperscript{*}Numerical data points: \href{https://tudatalib.ulb.tu-darmstadt.de/bitstream/handle/tudatalib/3625.3/figure_6.json}{https://tudatalib.ulb.tu-darmstadt.de/bitstream/handle/tudatalib/3625.3/figure\_6.json}
\label{fig:total_costs}
\end{figure}

All novel control methods presented were operated with the same 30-minute scenario, cf. Figure~\ref{fig:use_case}\,(c)\,(i).
The total resulting costs -- normalized to the minimum costs of all approaches -- are shown in Figure~\ref{fig:total_costs}.

The central MPC has the lowest costs, which are used as a normalization factor.
The highest costs of \SI{140}{\percent} are observed for the central MADRL - the costs thus vary by up to \SI{40}{\percent}.
As expected, MPC is better than DMPC, although the difference is small.
Partially observable MADRL, on the other hand, is unexpectedly better than central MADRL, even though the latter actually has more information available.
An interpretation of this communication and information influence will follow later.

\begin{figure}[h]
\includegraphics[width=13 cm, trim={0 7.85cm 6.9cm 0cm},clip]{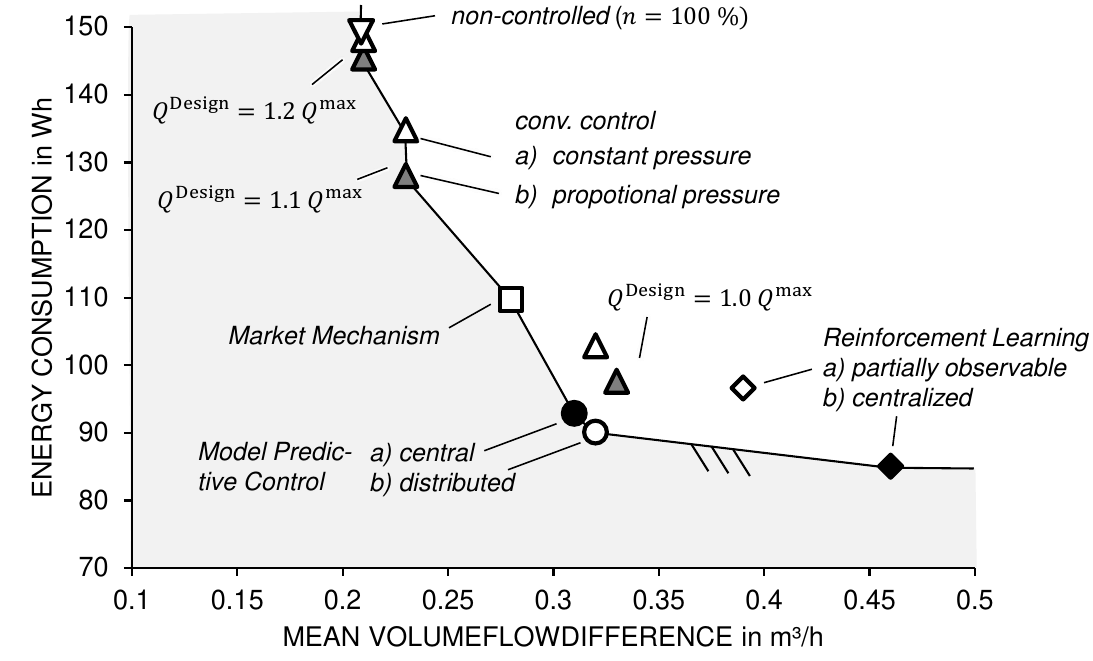} 
\caption{Trade-off between volume flow deviation and energy consumption.
         The standard deviation of the volume flow deviation along the load profile is between \SI{0.31}{\cubic \meter \per \hour }  (MPC) and
                                                                                               \SI{0.43}{\cubic \meter \per \hour } (Market Mechanism).\textsuperscript{*}}
\scriptsize\raggedright\textsuperscript{*}Numerical data points: \href{https://tudatalib.ulb.tu-darmstadt.de/bitstream/handle/tudatalib/3625.3/figure_7.json}{https://tudatalib.ulb.tu-darmstadt.de/bitstream/handle/tudatalib/3625.3/figure\_7.json}
\label{fig:tradeoff}
\end{figure}

In the control objective, both the deviation from the desired volume flow and the energy consumption are taken into account in a joint function.
The individual components and the trade-off between them are shown for the different approaches in Figure \ref{fig:tradeoff}.

There is a clear trade-off between volume flow deviation and energy consumption among the different approaches.
Each of the novel approaches, except one, are best with respect to some weighting of the objectives, i.e. the approaches show a Pareto front.
Central MADRL has the lowest energy consumption, but the highest volume flow deviation.
The market mechanism is exactly the opposite.
Partially observable MADRL is dominated, i.e. it is worse than the MPC approaches with respect to both criteria.

The conventional strategy shows that the constant pressure control has a higher energy consumption compared to the proportional pressure control, but a lower volume flow deviation.
This can also be observed by increasing the safety factor.
Without a safety factor, the approaches are dominated by MPC and market mechanism.
For higher safety factors, the conventional strategies show the lowest achievable volume flow deviation -- but a significantly higher energy consumption.
The uncontrolled case represents an upper limit for the lowest possible volume flow deviations, which, however, leads to very high energy consumption.

In all approaches, the preference in the trade-off can be controlled.
In MPC and MADRL this is done by the factors in the objective function, in the market mechanism by the allocation of costs and budgets.
However, despite the same weighting factors in MPC and MADRL, they show a different result, which is also evident in the total costs analyzed earlier.

\begin{figure}[h]
\center
\includegraphics[width=\textwidth]{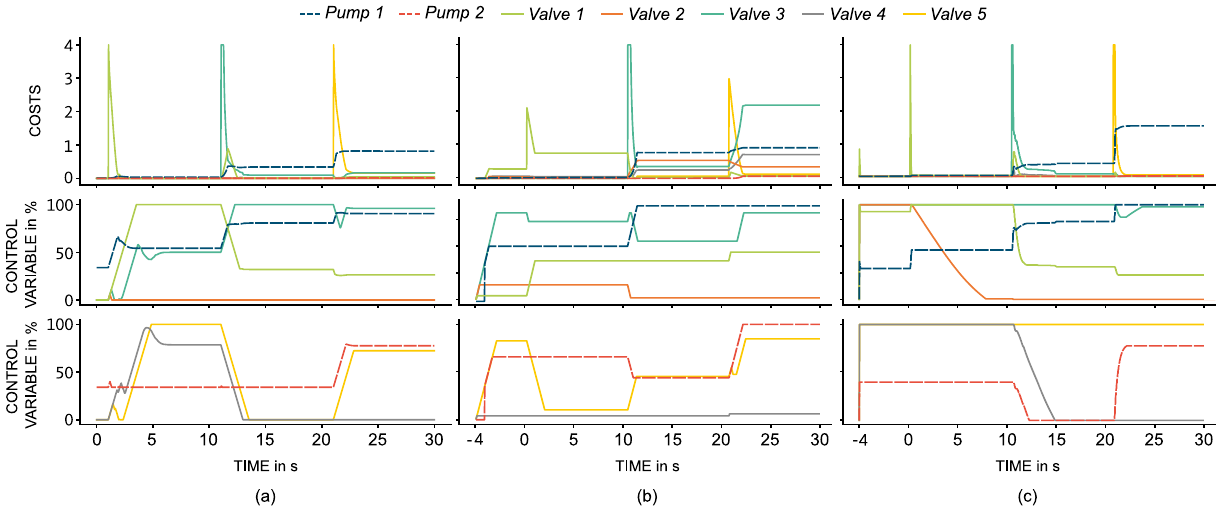}
\caption{Local costs (top) of the individual agents and control variables for lower subsystem (Pump 1 and Valves 1 -- 3) (middle) and upper subsystem (Pump 2 and Valves 4 and 5) (bottom) for the approaches (a) DMPC (b) MADRL partially observable and (c) Market Mechanism.\textsuperscript{*}}
\scriptsize\raggedright\textsuperscript{*}Numerical data points: \href{https://tudatalib.ulb.tu-darmstadt.de/bitstream/handle/tudatalib/3625.3/figure_8_values.json}{https://tudatalib.ulb.tu-darmstadt.de/bitstream/handle/tudatalib/3625.3/figure\_8\_values.json}
\label{fig:30sec}
\end{figure}

The demand profile shown in Figure~\ref{fig:use_case}\,(c)\,(ii) is  a simplified scenario, i.e. valves 1, 3 and 5 have an initiating demand one after the other.
The results shown in Figure \ref{fig:30sec} allow a closer look at the behavior of the individual agents and to check the comprehensibility.
Here, it needs to be stated that while the DMPC approach used the loadprofile presented in Figure~\ref{fig:use_case}\,(c)\,(ii), the other two approaches required a longer initial phase without demand.
Accordingly, a period of \SI{4}{s} without demand was added before the actual beginning of the load profile as indicated by negative values on the time axis in Figure~\ref{fig:30sec}.

In Figure \ref{fig:30sec} (a), the upper plot shows that the flow demand of the valves is quickly compensated and thus the cost of the valves decrease.
However, the costs of the pumps increase due to the higher energy consumption.
In the middle plot, it can be seen that the valves -- as expected -- open as soon as they have a demand.
After valve 3 has a demand (starting from second 11) valve 1 closes to avoid overfilling.
At some points, obviously energetically inefficient states can be identified.
For example, pump 2 (lower plot) never turns off, although it is not needed until second 21.
Furthermore, valve 5 is not fully open in the last third and thus the pump and valves work against each other.
A possible explanation for the pump not turning off is the local controller of this pump.
At this point there is a discontinuity (switch off from \SI{40}{\percent} speed to \SI{0}{\percent} speed), which is difficult to be taken into account by the local PID controller.
The incomplete valve opening may be due to a poor model or only locally optimal solution.

When using the MADRL approach (b), the behavior appears to be similar in general. The demands are quickly satisfied, but non-efficient states occur.
This is due to a suboptimal policy of the individual agents.
It is noticeable that in the last third, the performance of valve 3 is poor (high costs), which improves the performance of valves 1 and 5.
Thus, valve 3 supports the other valves.

In the market mechanism (c), the demands are also quickly satisfied.
Initially, pump 2 is also unnecessarily switched on, however, due to the negotiations, the actions become more comprehensible.
The data of the negotiations show, that pump 1 sells no volume flow to pump 2 at this time.
The fact that the pump is still not switched off can therefore only be due to the local controller.

\subsection{Influence of Available Information}
This section addresses research question (ii) concerning the available information from substitute models and communication between agents.

First, the effect of modelling effort is considered for the overall costs as well as functional quality and effort regarding energy consumption, cf. Figures~\ref{fig:total_costs} and \ref{fig:tradeoff}.
The model-based approaches MPC and DMPC have significantly lower costs. The approaches that use less model knowledge (MADRL and Market Mechanism) have higher costs. Thus, the tradeoff between reaching the objective and system knowledge -- which is associated with high modelling effort -- becomes clear.
MADRL uses data-driven models as a substitute for analytical model knowledge and achieves better energy efficiency, though at the cost of worse functional quality.
Market mechanism, on the other hand, uses no model for the system-wide behavior, yet achieves the best values among the distributed approaches for functional quality albeit with higher energy consumption.

\begin{figure}[h]
\center
\includegraphics[width=13 cm, trim={0 10cm 6cm 0cm},clip]{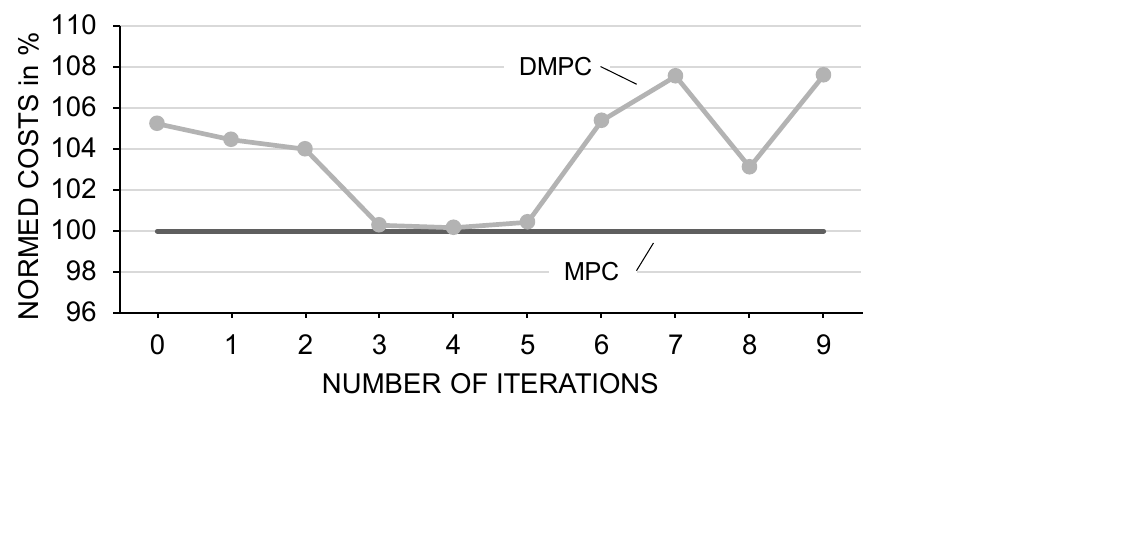} 
\caption{Costs of DMPC depending on number of communication rounds between agents against MPC benchmark.\textsuperscript{*}}
\scriptsize\raggedright\textsuperscript{*}Numerical data points: \href{https://tudatalib.ulb.tu-darmstadt.de/bitstream/handle/tudatalib/3625.3/figure_9.json}{https://tudatalib.ulb.tu-darmstadt.de/bitstream/handle/tudatalib/3625.3/figure\_9.json}
\label{fig:com}
\end{figure}

As presented in Section~\ref{sec:dmpc-meth}, in DMPC the agents may exchange information regarding the respective values of their manipulated variables and recalculate their optimal solution based on the updated information.
The influence of the number of such rounds of communication was investigated for DMPC, and the results can be found in Figure~\ref{fig:com}.
The overall costs for MPC were taken as a benchmark, against which the cost of DMPC with rounds of communication varied between \SI{0}{} and \SI{9}{}.
As expected, the costs for DMPC are consistently higher than for MPC.
As the number of communication rounds increases, the costs decrease until a minimum is reached at \SI{4}{} while a sudden drop in costs occurs between \SI{2}{} and \SI{3}{}.
With the number of communication rounds increased beyond \SI{4}{} the costs again increase which is contrary to expectations.
The costs for MPC should represent an asymptotic limit towards which the costs for DMPC would be expected to converge with increasing number of communication rounds.
The presented results suggest that rounds of communication beyond an optimal amount cause confusion rather than improved solutions.

As seen in Figure~\ref{fig:total_costs}, partially observable MADRL achieves lower total costs than central MADRL, despite being a dominated solution (cf. Figure~\ref{fig:tradeoff}).
Intuitively, central MADRL is expected to perform better, as universal availability of target values and control variables for each agent are closer to a centralized control.
A possible interpretation of these results is that the amount of information available to agents in the case of central MADRL causes information overload rather than helping coordination among agents.
Partially observable MADRL, on the other hand, achieves a better result with each agent only using its own variable values and all current demands in the system.
Thus, this approach seems to represent an adequate compromise, with necessary information shared between agents while avoiding distracting agents with surplus information.

Similar to the rounds of communication in the case of DMPC, rounds of bidding were investigated for market mechanism control.
This proved to have no influence on the overall costs of market mechanism control for the $30\ \mathrm{min}$ scenario.

To summarize, the presented results suggest that excess amounts of information can deteriorate the results of distributed control approaches.
Nevertheless, up to a certain point, increased sharing of information between agents improves the performance of distributed control.

\subsection{Performance under Disruption}
\begin{figure}[h]
\center
\includegraphics[width=13 cm, trim={0 8.8cm 6cm 0cm},clip]{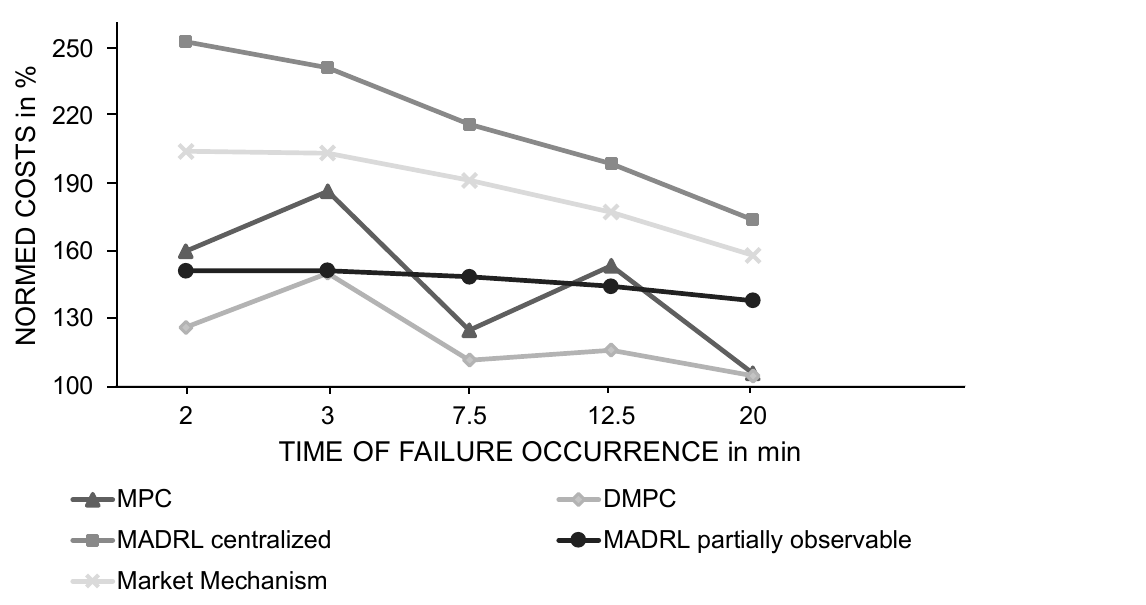} 
\caption{Comparison of costs from MADRL, (D)MPC, and market mechanism for various times of failure occurrence in the communication.\textsuperscript{*}}
\scriptsize\raggedright\textsuperscript{*}Numerical data points: \href{https://tudatalib.ulb.tu-darmstadt.de/bitstream/handle/tudatalib/3625.3/figure_10.json}{https://tudatalib.ulb.tu-darmstadt.de/bitstream/handle/tudatalib/3625.3/figure\_10.json}
\label{fig:resdisr}
\end{figure}

Finally, results for answering research question (iii) are presented in Figure~\ref{fig:resdisr}.
The disruption as described in Section~\ref{sec:disrmod} is triggered at different times during the \SI{30}{min} scenario.
The results for total costs for MPC, DMPC, central and partially observable MADRL as well as market mechanism in the disrupted case are related to the cost of the undisrupted case with MPC.
As expected, the cost in the disrupted case exceeds the cost in the undisrupted case for all control approaches.

The total cost for DMPC is lower than for MPC for every time of failure occurrence, i.e. DMPC allows for improved control when faced with the given disruption.
With later times of failure occurrence, the costs of both approaches converge at the time of low to no demand in the system.
The specific time of failure occurrence affects costs for both MPC and DMPC.
The overall trend shows lower increases of cost for later times of failure occurrence for both approaches.
This behavior is to be expected due to shorter periods of disruption as well as lower overall demand with increasing time.
However, when failure occurs at \SI{3}{min} and \SI{12.5}{min} higher costs are recorded for both approaches than at \SI{2}{min} and \SI{7.5}{min} respectively.
This is due to unfavorable demand distributions at the given times of failure occurrence, which are then propagated for the remainder of the \SI{30}{min} scenario.

As expected from the results shown in Figure~\ref{fig:total_costs}, the cost for central MADRL exceeds that of partially observable MADRL.
The costs for both approaches are increased compared to the undisrupted case, cf. Figure~\ref{fig:total_costs}, though the effect on partially observable MADRL is small.
The costs in the disrupted case of partially observable MADRL are of the same magnitude as those of MPC and DMPC under disruption.
A trend of lower increase in costs due to disruption with later times of failure occurrence is observable
especially for central MADRL while the effect is much less pronounced for partially observable MADRL.
The trend is not broken by the effect described for MPC and DMPC.
This results in costs for partially observable MADRL being as low as those of DMPC or lower than those of MPC for certain times of failure occurrence.

The performance of market mechanism is also severely affected in the case of disruption and the costs exceed those of MPC, DMPC and partially observable MADRL.
With later times of failure occurrence, the increase in costs is reduced, similar to central MADRL.

To summarize, answering research question (iii),
it is concluded that disruptions affect centralized control represented by MPC to a greater extent than distributed control as represented by DMPC and partially observable MADRL
while the performance of central MADRL and market mechanism deteriorates significantly beyond the range of the other approaches.
Among the distributed approaches, DMPC and partially observable MADRL are best suited to cope with disruptions to the communication infrastructure,
while market mechanism appears to be more vulnerable.

\section{Discussion}
\label{sec:disc}

Moving beyond central, optimal control of fluid systems, the presented work investigated three approaches to distributed control, DMPC, MADRL, and market mechanism, stemming from the domains of control theory, machine learning, and game theory respectively.
These approaches were implemented and tested using a generic fluid system model for water supply and a randomly generated demand pattern based on a typical demand profile for a building.
Corresponding to the challenges presented in Section~\ref{sec:intro}, the performance of the approaches is evaluated regarding functionality, effort in terms of energy consumption and modelling, availability and acceptability.

Converting deviation from control objective as well as consumed energy to standardized costs allows assessing functional quality, effort in terms of energy, and availability quantitatively, whereas the criteria modelling effort and acceptability are assessed qualitatively based on the experience of implementing the approaches.
The quantitative results have been presented in detail in the preceding section.

MPC and DMPC find the best trade-off between control objective and energy consumption, as both values are part of the objective function in the implemented optimization problem. In accordance with the first hypothesis, the distributed approach of market mechanism requires higher amounts of energy whereas central MADRL, by contrast, is more efficient in terms of energy consumption, yet this comes at the cost of significantly worse performance regarding functional quality.

Implementation efforts for MADRL and MPC/DMPC are extensive and since the substitute model for MPC/DMPC is generated using an artificial neural network, both approaches are placed towards the black box side of the white box vs. black box spectrum, implying less transparency.
Modification of the fluid system entails further effort for both approaches since the substitute model needs to be relearned from scratch, impeding flexibility and scalability of the approaches.
By contrast, market mechanism functions based on simple comprehensible rules and algorithms, rendering it a transparent approach.
System modifications are easily integrated, as the overall communication structure is maintained and the governing rules apply to all market participants.
This shows advantages with regard to flexibility and scalability of the approach.

Regarding the hypothesis concerned with availability and reliability in the face of failures in the communication, the results support the conclusion that distributed approaches are more reliable than centralized approaches with the exception of market mechanism.
The high information exchange requirements of central MADRL are equally detrimental to performance under disruption as the necessity to gather all information centrally in the case of MPC.
However, before arriving at a conclusive statement regarding the hypothesis, methods from information and network theory need to be applied to further investigate the hypothesis.

Existing studies have shown the feasibility of agent-based control of transport systems in general and fluid and related systems such as HVAC and water distribution systems in particular.
Most studies use rule-based mechanisms for designing agents and interaction, while other studies use approaches from the domain of machine learning.
Few compare results to benchmarks set by control using optimization methods or conventional control strategies.
The presented study considered distributed control approaches from three different domains, all based on multi-agent systems, and applied them to a generic fluid system.
A systematic assessment of the approaches allowed to compare them to one another as well as to benchmarks set by conventional control strategies and centralized control using global optimization methods.

Advantages and drawbacks of the three approaches have been identified, yet all three approaches were shown to be feasible in simulation.
Transferring approaches from the world of numerics to physical systems poses great challenges with regard to uncertainty, delay tolerance and system inertia.
In this study, only the reliability in the face of failures in the communication network was considered.
Failures may also occur in the technical system which constitutes the environment of the agent system.
The reliability of agent-based control against such disruptions should be investigated in future work.

Turning from physical applications to refinement of the methods considered in the presented work, there is great potential for further work.
DMPC may be improved by studying various solvers and optimizing parameters such as prediction horizon and control variable boundaries to ensure global optimum operation points at all times.
Looking at MADRL, hyperparameter optimization may be a first step towards improving the performance.
In order to address the aspect of flexibility and scalability, transfer learning could be considered for training basic agents which are then inserted into existing systems after which the entire system would adapt, learning an updated system model which includes the new components.
System safety is a further concern, with methods required to be included in the training process that prohibit unsafe operating conditions.
The market mechanism investigated here was devised using heuristic, ad-hoc rules which were only modelled for the components as far as necessitated by the use-case.
Aside from generalising the algorithms for further system configurations, methods for optimizing mechanism design could be closely studied to ensure optimal and fair allocation of goods.
Furthermore, game theoretic approaches beyond markets may provide further insights for devising appropriate mechanisms.

Applications of agent-controlled fluid systems may be in bounded environments, such as factory cooling circuits or domestic heating systems.
However, water distribution systems, waste-water disposal systems, gas and district heating are large-scale infrastructures that may also be considered for agent-based control.
These infrastructures provide vital services to communities, which raises two further aspects.
Firstly, the system boundary of the environment of the agent-system would need to be expanded to include the socio-technical system of consumers using the technical system controlled by agents.
This will render demand patterns at valves more complex.
Secondly, with the vital importance of these infrastructure systems for communities served by them, questions of resilience need to considered in the design approaches in order to provide minimal functionality and the possibility of recovery when critical adverse events occur.

\section{Conclusions}

The presented work studied three distributed agent-based control approaches stemming from the domains of game theory, control theory and machine learning respectively in application to a generic fluid system with the medium water.
The performance of the approaches was studied in simulation and assessed quantitatively and qualitatively, and compared to benchmarks of conventional control strategies and centrally computed optimal control.
Future research directions include application of the distributed agent-based control approaches to a physical fluid system and experimental validation, as well as refinement of the design and methods underlying each of the approaches.

\section*{Funding}
This work was co-funded by the \emph{Forschungsfond Pumpen} wihtin the \emph{Forschungsvereinigung Pumpen+Systeme}
of the \emph{Verband Deutscher Maschinen- und Anlagenbau} under the project \emph{Multiagentensysteme zur verteilten Regelung von Fluidsystemen}.
It has been co-funded by the LOEWE initiative (Hesse, Germany) within the \emph{emergenCITY center}.

\section*{Acknowledgements}
The authors would like to thank Haoze Yang, Daniele Inturri, Tobias Meck and Dennis K{\"o}nig for their work in the scope of the project.
We want to further thank all the participants of the working group \emph{Muliagentensysteme} of the \emph{Verband Deutscher Maschinen- und Anlagenbau, Pumpen + Systeme}
for the constructive and close collaboration and discussions.

\section*{Declaration of interests}
The authors declare that they have no known competing financial interests or personal relationships that could have appeared to influence the work reported in this paper.

\section*{Data availability statement}
Results data together with the numerical values presented in figures, the functional mock-up unit of the test rig model, and the load profile is available on the institutional repository under the following DOI: \href{https://doi.org/10.48328/tudatalib-984.3}{https://doi.org/10.48328/tudatalib-984.3}.
The Python code for running the simulations is available on demand.

\section*{Author Contributions}

Contributions following the CRediT taxonomy:

\textbf{Kevin T. Logan:} Conceptualization, Methodology, Validation, Data Curation, Writing -- original draft preparation, Writing -- review and editing, Visualization;

\textbf{J. Marius Stürmer:} Methodology, Software, Validation, Formal Analysis, Investigation, Data Curation, Writing -- original draft preparation, Writing -- review and editing, Visualization;

\textbf{Tim M. Müller:} Conceptualization, Methodology, Validation, Writing -- original draft preparation, Writing -- review and editing, Visualization, Project Administration, Funding acquisition;

\textbf{Peter F. Pelz:} Resources, Writing -- review and editing, Supervision, Project Administration, Funding acquisition.

\bibliographystyle{elsarticle-num}
\bibliography{bibliography.bib}

\end{document}